\documentclass[11pt,twoside,letterpaper]{article} 

\usepackage{microtype}
\usepackage[latin1]{inputenc}

\usepackage[hmarginratio=1:1,top=32mm,columnsep=20pt]{geometry}
\usepackage[labelfont=bf,textfont=it,font=small]{caption}
\usepackage{paralist}

\usepackage[pdftex]{graphicx}

\usepackage[rightcaption]{sidecap}
\sidecaptionvpos{figure}{t}

\usepackage{authblk}
\usepackage{geometry}
\usepackage{lscape}
\usepackage{amsfonts,fancyhdr,amssymb}
\usepackage{booktabs,multirow}

\usepackage{amsmath}
\newcommand{\argmin}{\arg\!\min}

\renewcommand{\thetable}{\arabic{table}}
\renewcommand{\thefigure}{\arabic{figure}}
\renewcommand{\theequation}{\arabic{equation}}
\renewcommand{\thesection}{\arabic{section}}

\usepackage{nameref}

\usepackage{hyperref}

\usepackage{cite}

\title{Fundamental structures of dynamic social networks}

\author[1]{Vedran Sekara}
\author[1,2]{Arkadiusz Stopczynski}
\author[1,3]{Sune Lehmann}
\affil[1]{Department of Applied Mathematics and Computer Science, Technical University of Denmark, Kgs. Lyngby, Denmark}
\affil[2]{Media Lab, Massachusetts Institute of Technology, Cambridge, MA, USA}
\affil[3]{The Niels Bohr Institute, University of Copenhagen, Copenhagen, Denmark}

\begin{document}

\maketitle

\begin{abstract}
\noindent Social systems are in a constant state of flux with dynamics spanning from minute-by-minute changes to patterns present on the timescale of years.
Accurate models of social dynamics are important for understanding spreading of influence or diseases, formation of friendships, and the productivity of teams.
While there has been much progress on understanding complex networks over the past decade, little is known about the regularities governing the micro-dynamics of social networks.
Here we explore the dynamic social network of a densely-connected population of approximately 1\,000 individuals and their interactions in the network of real-world person-to-person proximity measured via Bluetooth, as well as their telecommunication networks, online social media contacts, geo-location, and demographic data.
These high-resolution data allow us to observe social groups directly, rendering community detection unnecessary.
Starting from 5-minute time slices we uncover dynamic social structures expressed on multiple timescales.
On the hourly timescale, we find that gatherings are fluid, with members coming and going, but organized via a stable core of individuals.
Each core represents a social context.
Cores exhibit a pattern of recurring meetings across weeks and months, each with varying degrees of regularity.
Taken together, these findings provide a powerful simplification of the social network, where cores represent fundamental structures expressed with strong temporal and spatial regularity. 
Using this framework, we explore the complex interplay between social and geospatial behavior, documenting how the formation of cores are preceded by coordination behavior in the communication networks, and demonstrating that social behavior can be predicted with high precision.
\end{abstract}

\newpage

\section*{Introduction}
Human societies, their organizations, and communities give rise to complex social dynamics that are challenging to understand, describe, and predict.
Recently, network science has provided a powerful mathematical framework for describing the structure and dynamics of social systems~\cite{easley2010networks, newman2010networks,wasserman1994social}.
With deep roots in traditional sociology~\cite{simmel1950quantitative,goffman2005interaction}, a central challenge in the description of social systems is understanding social group behavior.
Using empirical data, such groups (or communities) have recently been shown to be highly overlapping and organized in a hierarchical manner~\cite{palla2005uncovering, palla2007quantifying, clauset2008hierarchical, ahn2010link}.
Without an understanding of the fundamental meso-level structures and regularities governing social systems, modeling and predicting behavioral patterns remains a challenge~\cite{holme2012temporal,starnini2013modeling}.

While a coherent mathematical framework is not yet in place, existing research suggests that social dynamics are far from random.
For example, the existence of strong regularities for individuals in human populations has been well documented within mobility patterns~\cite{gonzalez2008understanding, eagle2009eigenbehaviors,song2010limits,lu2012predictability}, and in social systems pair-wise interactions show clear patterns occurring at multiple timescales from seconds to months~\cite{saramaki2015seconds}.
For groups of interacting individuals, however, an understanding of the fundamental structures and their temporal evolution across timescales has proven elusive so far, suggesting a potential for a better understanding and models describing important processes such as spreading of influence or diseases, formation of friendships, or productivity of teams.

Our work is based on a longitudinal (36 months) high-resolution dataset describing a densely-connected population of approximately 1\,000 freshman students at a large European university~\cite{stopczynski2014measuring}.
We consider interactions in the network of physical proximity measured via Bluetooth (see methods), complemented with information from telecommunication networks (phone calls and text messages), online social media (Facebook interactions), as well as geo-location and demographic data.

Until this point, community detection in dynamic networks has required complex mathematical heuristics~\cite{mucha2010community, gauvin2014detecting}.
Here we show that with high-resolution data describing social interactions, community detection is unnecessary. 
When single time slices are shorter than the rate at which social gatherings change, communities of individuals can be observed directly and with little ambiguity (Fig.\,\ref{fig:fig1}a).
Using a simple matching between time slices we can infer temporal communities. 
These dynamic communities offer a powerful simplification of the complex system of social interactions as it develops over time.

Below we describe these findings in detail. 
We then study the properties of gatherings and cores, show how appearances of social cores are preceded by increased communication among their members, and examine how cores tend to behave as social units. 
As an application of the framework, we show that the social dynamics of this population are highly predictable. 
 
\section*{Results}
Social interactions unfold on many timescales, with structures and regularities spanning from minute-by-minute changes to yearly rhythms and beyond, as observed in telecommunication networks~\cite{saramaki2015seconds} or online social networks~\cite{golder2011diurnal}.
Despite these inherent dynamics, most of the current understanding of social networks comes from the study of static network topologies, not including temporal aspects~\cite{easley2010networks, newman2010networks}.
Human social communities are inherently overlapping~\cite{palla2005uncovering, palla2007quantifying, ahn2010link, rosvall2014memory, evans2009line} with each and every individual participating in multiple communities. 
In time-aggregated networks, this structural property results in communities with more outgoing than incoming edges, the `hairball' structure visualized in Fig.\,1a (green network).
With minute-by-minute observation intervals of this network, individuals are constrained to a single community per time slice, and we can observe the basic structural elements directly and with little ambiguity, as illustrated in Fig.\,\ref{fig:fig1}a (blue network).
More generally, when time slices have a duration which is shorter than the rate with which the group composition changes, gatherings of individuals can be observed directly, rendering traditional community detection algorithms redundant (the short time-scale time slices, Fig.\,\ref{fig:fig1}a (blue network) are strikingly different from random subsamples of equal size from the daily networks, see SI). Matching connected components across time slices, we can study the temporal development of social \emph{gatherings}.
In our dataset, gatherings are typically together for a few hours (full statistics available in SI), and can be thought of as instances of social structures (\emph{cores}) that occur repeatedly over days, weeks, and months.

\subsection*{Gatherings}
In the physical proximity network, meetings are require all members to be present at the same time, and can exist between multiple individuals. 
This implies that gatherings can be directly identified as graph components (consisting of individuals in close physical proximity) within each 5-minute time slice, 
Therefore, we are able to use a simple hierarchical clustering to match groups across time slices (see SI), using a matching strategy is similar to Ref.~\cite{greene2010tracking}. 

\begin{figure}
\centerline{\includegraphics[width=\linewidth]{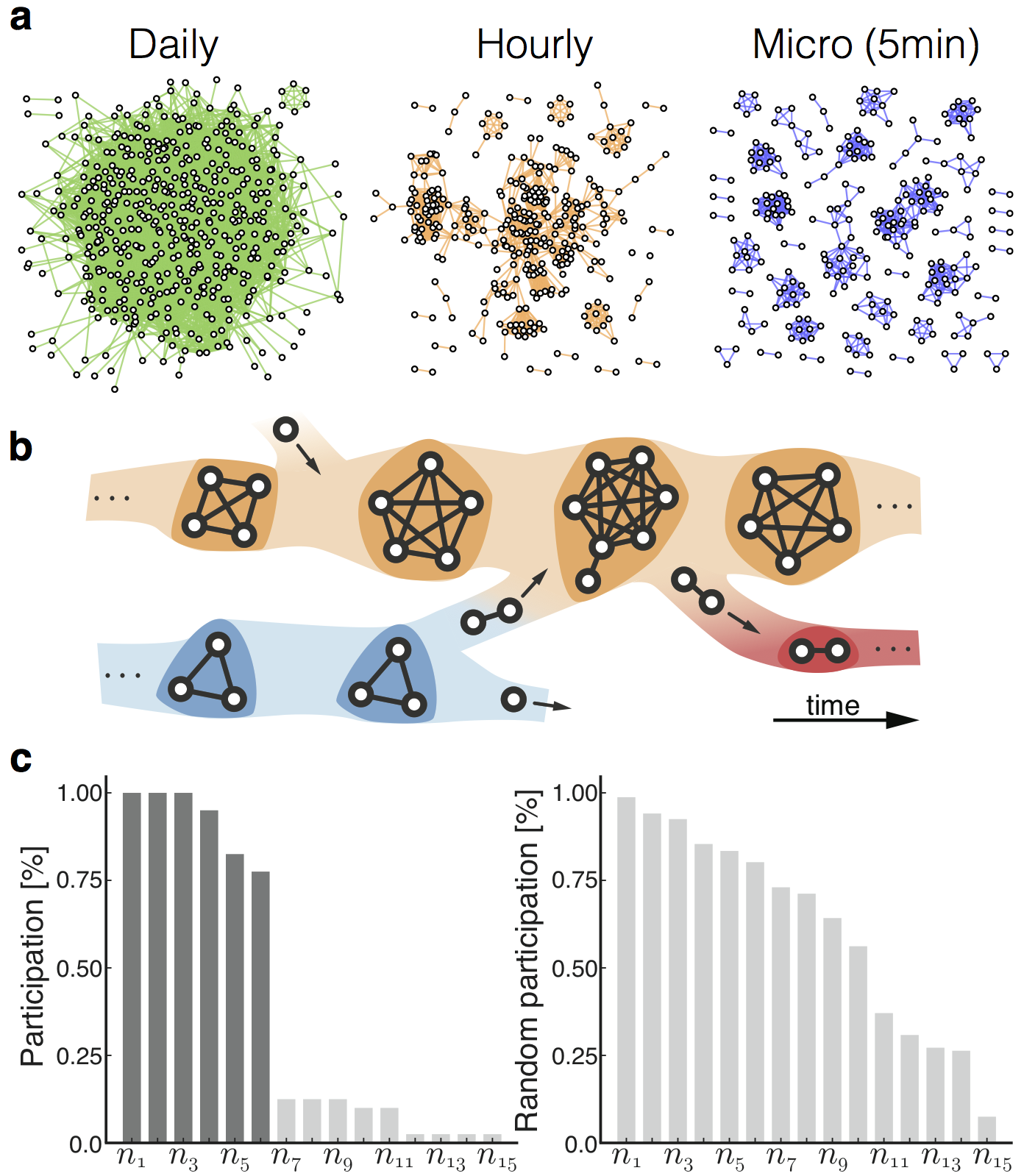}}
\caption{\textbf{Properties of gatherings.} \textbf{a}, The social network at different timescales. The network formed by face-to-face meetings within one day (green), $60$-minute (orange), and $5$-minute temporal aggregation (blue).
In the 5-minute time-slices groups are directly observable without much ambiguity, but the overlap between groups increases as time is aggregated across bins.
\textbf{b}, Illustration of gathering dynamics. 
Gatherings evolve gradually with members flowing in and out of social contexts.  \textbf{c}, Extracting cores from participation profiles. Dark gray bars denote nodes with participation levels above the maximal gap. Ordered participation profile for a empirical data (left), as a null model we use participation profiles generated from a uniform random distribution (right).\label{fig:fig1}}
\end{figure}

A node may be part of only a single gathering per time step, but over time individuals flow in and out of social gatherings, shifting their affiliation, and forming new gatherings as illustrated in~Fig.\,\ref{fig:fig1}b.
The gatherings display broad distributions in both size and duration, capturing meetings ranging from small cliques to large aggregations, and from short interactions on the order of minutes to prolonged encounters lasting many hours, covering a wide range of meeting types.
Gatherings are defined for any number of nodes greater than one, but since we are interested in group dynamics, we only discuss gatherings of size three or greater in the statistics reported here.
Since our cohort consists of university students, an important inhomogeneity in the data is between `work' activities that take place on campus (including scheduled classes) and `recreational' activities that take place off campus. 
Utilizing GPS information, we find that $42\%$ of gathering take place on-campus (work) and $58\%$ take place off-campus (recreation).
Comparing work/recreation statistics, recreational gatherings tend to be smaller but last considerably longer, illustrating that the context of meetings can influence their properties (see SI for full statistics).

The fluid behavior illustrated in Fig.\,\ref{fig:fig1}b results in `soft' gathering boundaries, with some members participating for the total duration of the gathering and others participating only briefly.
Peripheral members can be acquaintances briefly interacting with members of the social core, but can also be spurious connections in the data: nearby strangers eliciting a Bluetooth measurement not corresponding to a social connection.
In spite of these soft boundaries, we find that gatherings are characterized by a stable core of individuals that are present during the majority of each meeting (see SI).
This stable core is expressed in participation profiles as illustrated in Fig.\,\ref{fig:fig1}c.
Here, a participation profile is the sorted fraction of time each node has participated in a particular social context, normalized by its total lifetime. 
A pronounced core structure implies a gap in the participation profile, separating the core members (Fig.\,\ref{fig:fig1}c, left, dark gray bars) from the peripheral nodes  (Fig.\,\ref{fig:fig1}c, left, light gray bars).
We test if the gap is statistically significant by comparing to an ensemble of profiles generated by a random process (Fig.\,\ref{fig:fig1}c, right), where a random participation level between $0$ and $1$ is assigned to each node from a uniform distribution.
Based on the ensemble, we estimate the average expected gap size and deviation.
If the gap observed in the empirical data is greater than the average null-model gap $\mu_{\text{random}}$ plus one standard deviation $\sigma_{\text{random}}$, we accept the core as significant. 
According to this criterion, we find that $7\,146$ out of the $7\,320$ ($97.6\%$) inferred communities display a pronounced core structure.

\subsection*{Cores}
A gathering represents the time evolution of a single meeting between a group of individuals. 
In most cases, however, gatherings are an instance of a lasting social context (e.g.~a group of friends or class-mates), and we observe the same nodes participating in subsequent gatherings occurring repeatedly over the following days, weeks, and months. 
We call the social structures corresponding to all gatherings of the same set of individuals \emph{cores}.
We argue below that cores represent a fundamental structure expressed in dynamic social networks with strong temporal and spatial regularity. 
Below we discuss cores. 
How cores are identified, and how cores can be used to quantify the regularity of social interactions, as well as predict future behavior of individuals in our dataset.

\begin{figure}
\centerline{\includegraphics[width=\linewidth]{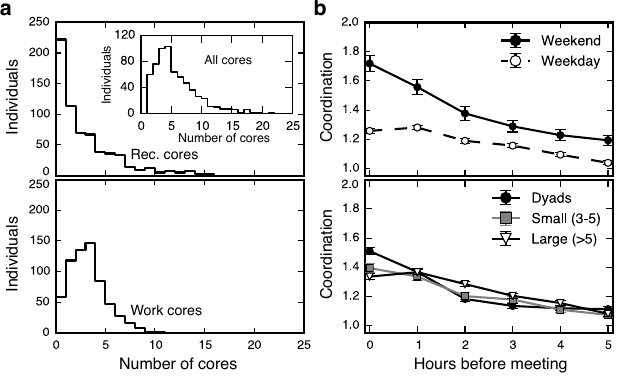}}
\caption{\textbf{Cores summarize social contexts for individuals.} 
\textbf{a}, The distributions of work and recreational core memberships, the top panel shows that individuals typically participate in only one or two recreational contexts, although the tail of the distribution contains individuals with more gregarious behavior. 
Inset shows the distribution of participation in cores overall (both work and recreation). 
\textbf{b}, Coordination prior to meetings defined as increase of phone calls and messages occurring within a hourly time-bins prior to the meeting relative to a null model based on the average hourly telecommunication behavior for each participant. 
More coordination is required to organize meetings during weekends than during weekdays. 
Larger meetings do not require additional coordination per participant. \label{fig:fig3}}
\end{figure}

In this population, the number of appearances per core is a heavy-tailed distribution; most cores appear only a few times, while the most active cores can appear multiple times per day over the full observation period (see SI).
Here we focus on the temporal patterns of recurring gatherings, restricting our analysis to cores of size three or greater that are observed more than once per month, on average.
The split between activities that occur on campus is important for cores as well as gatherings, and Fig.\,\ref{fig:fig3}a shows the difference of how individuals engage and spend time in different social contexts (\emph{work cores}---primarily observed on campus---and \emph{recreational cores}---primarily observed elsewhere). 
While the distribution of number of recreational cores per person is broad, most participants are part of only one or two recreational cores (Fig.\,\ref{fig:fig3}a, top panel). 
The distribution of work cores per person is localized with an average of $2.74\pm 1.85$ work cores per person (presumably corresponding to classes and study groups).

We find that cores leave traces in other data channels that emphasize the differences between work and recreational meetings. 
One such trace is coordination behavior, which we can explore by investigating how call and text-message activity increases in the time leading up to a meeting.
We define a core's level of coordination $c_t$ at time $t$ as the average increase of activity of its members prior to a meeting. 
Specifically, we set $c_t = 1 / N  \sum_{n=1}^N a_t^n / \widetilde{a_t^n}$, where $N$ is the number of participants, $a_t^n$ is the individual activity of person $n$ in time-bin $t$ (indexing the hours-of-the-week), compared to an individual baseline which is simply the person's average activity $\widetilde{a_t^n}$ in that hour of the week; to generate the curves in Fig.\,\ref{fig:fig3} we then average over cores.
We observe clear evidence of coordination prior to meetings, with a stronger effect during weekends (Fig.\,\ref{fig:fig3}b, upper panel). 
This result explicitly shows the extra coordination cost associated with non-schedule driven interactions.
Conversely, the coordination cost per participant does not depend on the size of the gathering in question (Fig.\,\ref{fig:fig3}b, lower panel), suggesting a social optimization process takes place (since the number of social connections grows as $n(n-1)/2$ for a group of $n$ participants).

\subsection*{Modeling the social network}
In order to place the significance of the basic statistics regarding gatherings and cores in perspective, we create a null model for the full network dynamics (see SI for full details).
We model each time-slice as a random geometric graph (RGG), and generate dynamics by representing each node as a random walker, similar to models known from the literature \cite{starnini2013modeling}.
For reasonable parameter values, we find that this \emph{dynamic RGG model} creates gathering-like components, and is able to recreate qualitative features of the distribution of both gathering sizes and lifetimes. 
Thus, as suggested in the literature \cite{starnini2013modeling}, we are in fact able to model key statistical features of gathering-occurrences across a single day, using a very simple model.

The fundamental difference between our empirical system and the dynamic RGG model arises because the model does not capture correlations between groups of nodes. 
This means that the dynamic RGG model does not generate recurring gatherings, and is unable to model dynamics across weeks and months. 
The fact that gatherings in the dynamic RGG model only occur once means that empirical statistics on the level of cores are not reproduced by the model.

\subsection*{Cores are social units}
It is well established that two individuals who share a social tie have correlated mobility patterns~\cite{crandall2010inferring, wang2011human, de2013interdependence,cho2011friendship}.
Because interactions with social ties tend to happen at irregular intervals~\cite{miritello2013temporal}, using this correlation for location prediction has proven difficult.
Expressed differently, we know that the location of person $A$ is going to be predictive of person $B$'s location (and \emph{vice versa}) when they meet, but we rarely know at which point in time $A$ and $B$ will meet. 

As we have defined them, cores represent meetings between $n>2$ individuals, characterized by the fact that all $n$ members are usually present when the core is active. 
Therefore an observation of an incomplete set of core members implies that the remaining members will arrive shortly, a fact which can be used for prediction. 
We call this the property of cores the `social unit' attribute.

We illustrate the social unit attribute using cores of size three.
Given that two members of a core are observed, we measure the probability that the remaining member will arrive within one hour.
To avoid testing on scheduled meetings, we only consider weekends and weekday evenings and nights (6pm-8am), where meetings are not driven by an academic schedule.
Further, we test on a month of data that has not been used for identifying cores.
We compare the behavior of actual cores to two null models, both based on an aggregated graph of daily interactions. 

The first null model (random), simply illustrates that our results are not driven by spurious co-locations.
In the random null model, we create an aggregated daily graph for each day in the observation period.
We then choose a random day and pick three individuals at random. 
If two of those individuals appear in the same location, we measure the probability of the third node arriving within one hour (including the reference group in the statistics only if at least two of its nodes are co-located during the day, see SI).
This situation is illustrated using blue nodes in Fig.\,\ref{fig:fig4}a.
The random null model shows no predictive power (Fig.\,\ref{fig:fig4}b).

For the second null model (BFS) we test the effect of pairwise friendships within groups of three. 
Here, we form reference groups using a breadth first search (BFS) strategy.
We choose a random day and based on a randomly chosen seed node, we perform BFS steps until the neighborhood-set is large enough to form a group of size three; then we create a reference group based on the seed and two random nodes from the neighborhood. 
This situation is illustrated in Fig.\,\ref{fig:fig4}a, where the BFS starts from a yellow seed node expanding to two randomly chosen neighbors marked in green.
We then choose situations where two nodes from this set are subsequently co-located and test how often the third BFS reference-group member arrives within the next 60 minutes (see SI).
The BFS model implies at least pairwise relationships between the (yellow) seed node and the the two other (green) group members.

Comparing the performance of cores with the two null models (Fig.\,\ref{fig:fig4}b) we find support for the social unit attribute. 
The random null model clearly shows that spurious connections do not carry a signal, and the BFS model illustrates that even if all three nodes form a connected subgraph, those subgraphs have a less than 10\% probability of the third member arriving within the hour. 
In contrast, the cores capture arrivals of the remaining member in close to $50\%$ of cases in spite of cores having soft boundaries.
These observations provide support for, and quantify, the social unit attribute. 
Cores require all members to be present.

\subsection*{Ego perspective: Core participation forms social trajectories}
\begin{figure}
\centerline{\includegraphics[width=0.6\linewidth]{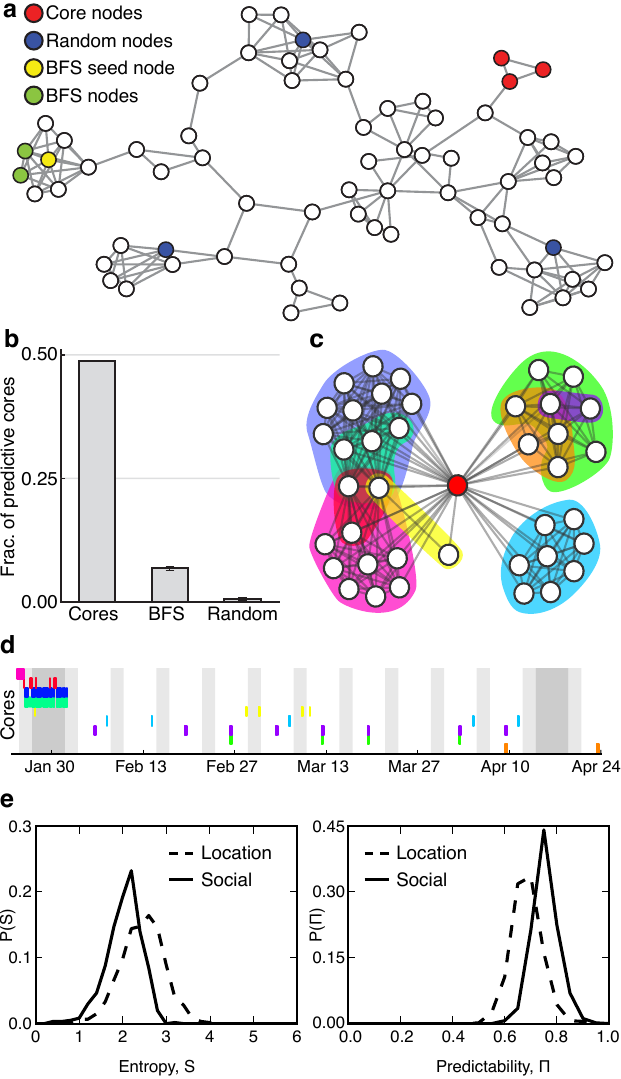}}
\caption{\textbf{Cores predict social behavior.} \textbf{a} Example of an aggregated graph of daily interactions, illustrating cores and construction of null models. Blue nodes correspond to the random null model, yellow/green nodes illustrate the BFS null model, and red nodes show a three-core. See main text for full details. \textbf{b} Percentage of predictive groups in each category: Cores, BFS reference-groups, and random groups. Reference model errorbars are calculated across n = 100 independent trials. 
\textbf{c}, Example ego view of communities; we observe overlapping and nested structures. \textbf{d}, The temporal instantiations of the cores in \textbf{c}. Time runs horizontally and each row corresponds to cores (colors matching \textbf{c}).
\textbf{e}, The distributions of time-correlated entropy ($S$) and predictability ($\Pi$) for social and location patterns. We find that overall social patterns tend towards lower entropy than geospatial traces, resulting in higher predictability. 
\label{fig:fig4}}
\end{figure}

From the perspective of an individual, we find the situtation displayed in Fig.\,\ref{fig:fig4}c, where each ego is involved in a number of overlapping and nested cores; the distribution (number of memberships) is shown in Fig\,\ref{fig:fig3}a, inset.
Figure~\ref{fig:fig4}d shows when cores are activated across our observation period from ultimo January to ultimo April, with time running horizontally and each row/color corresponding to a core, with colors matching Fig.\,\ref{fig:fig4}c; we call the core instantiation profile an individual's `social trajectory'.

Replacing the complex temporal dynamics of the social network with the sequence of participations in the core instantiation profile offers a powerful simplification of the complexity of a dynamic social network.
Since each core represents a social context the set of an individual's cores provides a finite set of states (a `vocabulary') for quantifying their social life.

The social trajectory shown in Fig.\,\ref{fig:fig4}d also provides an important connection to research within human mobility. 
Based on mobility data, it has recently been shown that human mobility patterns are regular, and that given a sequence of location-observations, a person's geographical location in the next time-bin can be predicted with high accuracy (an average of 93\% of the time)~\cite{song2010limits}.
Interestingly, the problem of predicting the next instantiation of a social core is equivalent to predicting the next location in a sequence of locations, once we summarize the dynamic network of social interactions using the social trajectory formulation.
In the analogy, a social core corresponds to a location in physical space, and instantiating a gathering corresponds to visiting that location.
This implies that we can use the methods developed in~\cite{song2010limits} to estimate an upper bound for the predictability of social trajectories.
The upper limit of predictability is derived from the amount of repetition (routine) encoded in a sequence of observations. 

The central property needed to estimate the bound on predictability is the time-correlated entropy $S$, which quantifies the amount of uncertainty within a data sequence, accounting for frequency and ordering of states. 
The upper bound on predictability, $\Pi$, can then be determined by applying a limiting case of Fano's inequality~\cite{fano1968transmission} (see SI).

Since we have access to participants' location traces as well as their networks, we can calculate the upper bound on predictability for \emph{both} social trajectories and location traces across the population.
In order to incorporate the full complexity of social interactions in the calculation of the time-correlated entropy, we include cores with any number of appearances, as well as cores consisting of two nodes in these calculations.
In Fig.\,\ref{fig:fig4}e, we display the distributions of time-correlated entropy and corresponding upper bound on predictability for the population. 
Comparing the two distributions, we find that the typical social behavior is characterized by lower entropy and thus higher predictability than the typical mobility behavior. 
In the dynamic RGG model, which we use as a null model for the full network dynamics, the temporal entropy does not converge; this implies that there is no social predictability in this model (see SI).

Comparing the empirical mobility behavior to the literature \cite{song2010limits,lu2013approaching}, we find a lower average predictability limit of approximately $80\%$ (Fig.\,\ref{fig:fig4}a).
The fact that the location-predictability is somewhat lower than the 93\% previously reported~\cite{song2010limits} is connected to a number of factors: the non-representative socio-demographics of the study population may play a role, the GPS location data used here has significantly higher spatial resolution than the cell towers used in previous work~\cite{song2010limits,lu2013approaching}---a fact which is known to decrease the estimated predictability~\cite{lin2012predictability}), and we focus on predicting the next location in a sequence of locations rather than predicting the person's location in the next time-bin---a more challenging prediction task (see SI). 

The overall level of social and geospatial predictability is not correlated for individuals ($p$-value $=0.85$, see SI).
In spite of this lack of correlation we know that the predictability encoded in social or geospatial trajectories is closely related to daily and weekly schedules~\cite{mcinerney2012exploring}.
A deeper exploration of the connection between social and geospatial behavior across the week provides an example of how the social trajectories enable a joint socio-spatial description of an entire population.

\begin{figure}
\centerline{\includegraphics[width=0.6\linewidth]{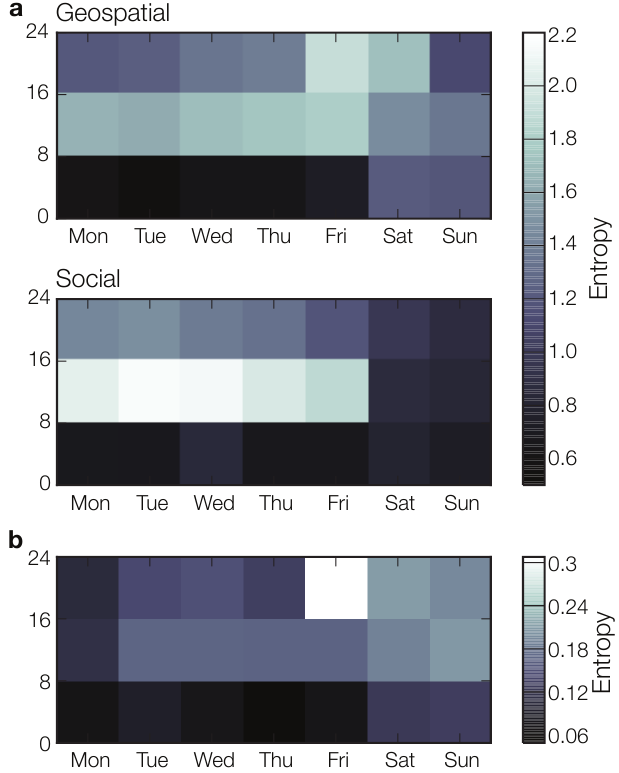}}
\caption{
\textbf{Temporal aspects of geospatial and social predictability.} 
\textbf{a}, Uncorrelated entropy respectively for location (upper) and social (lower) states, averaged across all individuals. Showing how unpredictable individuals are within 8-hour bins over the course of a week.
\textbf{b}, Geospatial entropy of cores, indicating when cores, i.e. individuals together in a social setting, explore new geospatial locations. 
\label{fig:fig5}}
\end{figure}

First we consider the typical behavior by calculating the uncorrelated entropy (see methods) for the social and geospatial trajectories independently. 
As a measure of disorder, the entropy quantifies how unpredictable a pattern is: the higher the entropy, the more users tend to distribute their time between many distinct states.
Figure~\ref{fig:fig5}a shows the entropy averaged over the population for each day of the week, in 8-hour bins. \emph{Nights} running from midnight to 8am, \emph{days} spanning 8am to 4pm and \emph{evenings} from 4pm to midnight.
Light colors indicate high entropy (complex behavior) and dark colors low entropy (simple patterns).

The geospatial behavior (Fig.\,\ref{fig:fig5}a, upper panel) is characterized by low entropy on weeknights, essentially corresponding to sleeping in one or two locations. 
The entropy is high during weekdays and assumes mid-range values during most evenings with a strong exception on Friday and Saturday nights, which are characterized by the highest entropy of the entire week.
These `party nights', are consistently the time-bins with highest average geospatial entropy, corresponding to exploration behavior.

The social behavior (Fig.\,\ref{fig:fig5}a, lower panel) is similar to the geospatial behavior across most of the week: Predictable nights, varied days, and evenings somewhere in between. 
On Friday and Saturday night, however, the social trajectories display behavior that is significantly different from the collective pattern arising from the geospatial trajectories.
In these time-bins, when the study participants are most exploratory in a geospatial sense, they appear to be highly conservative in social sense, displaying simpler and more predictable social behavior. 
This finding is consistent with Fig.\,\ref{fig:fig3}a, which shows that a large majority of participants focus their off-campus life on a small number (one or two) social cores. 

The observations above suggest that during the week, the population is characterized by a pattern of the same people meeting in the same places, with exploration peaking on Friday nights and weekends. 
On Friday night and weekends when the population is most unpredictable in their geospatial behavior, they are highly predictable in their social behavior, exploring a range of locations, but always with the same core of friends.

In interpreting this result, it is important to realize that we only observe the social interactions among participants of the experiment. 
While the geospatial data-stream is sampled evenly over the observation period, the social stream has a potential bias, e.g.~it is possible to go out with non-university friends on Fridays and weekends.
We can address this caveat by considering the behavior of \emph{cores} across the week (Fig.\,\ref{fig:fig5}b).
In Fig.\,\ref{fig:fig5}c we consider the behavior of cores across time, displaying the uncorrelated location entropy of `core location histories, i.e.~the sequence of locations visited by each core (averaged and binned as above).
In interpreting the absolute values of entropy in Fig.\,\ref{fig:fig5}b, note that cores typically only meet in a few locations; thus, in turn cores have fewer location states than individuals, resulting in smaller average values for the entropy. 

Figure\,\ref{fig:fig5} shows directly that cores do display a distinct exploratory behavior on Friday nights (and weekends more generally). 
The fact that geospatial exploration occurs as part of a social group, but constrained to certain time-bins reveals a complex interplay between time, location, and social context, and supports the hypothesis that at times when humans are most unpredictable in the geospatial domain, they display predictable social behavior.

\section*{Discussion}
The freshman participants of this study are not a representative sample of society as a whole, and we expect that some aspects of the findings presented here reflect that this is a particularly youthful demographic sample (e.g.~geospatial entropy peaking on Friday and weekend nights).
While the population studied here is not representative of society as a whole, we argue that the methods developed, as well as many of our findings do generalize to more representative populations. 
Below we summarize the most important of those findings.

It is often the case that the mathematical description of networks grows much more complex once the temporal dimension is included~\cite{holme2012temporal}, with even basic network properties, such as degree, clustering coefficient, or centrality having multiple competing generalizations to the temporal domain. 
Here, we observe the opposite effect, that additional temporal information can simplify the description
By observing the network at high temporal resolution we can directly observe gatherings, a phenomenon which is obscured in time-aggregated networks~\cite{ahn2010link}. 
In another context \cite{rosvall2014memory}, data on empirical network flow has been used as extra information in order to uncover overlapping communities.
A simple matching across time slices reveals dynamically changing gatherings with stable cores that provide a strong simplification of the social dynamics.
These cores leave traces in other data channels, for example eliciting coordination in the telecommunication network. 
The cores are fundamental units in the sense that they require all members to be present, a fact which we can use to predict the future location of other (non-present) members.

Connecting our findings to the literature of dynamic community detection, we note that many elegant methods exist in the literature that would allow us to detect gatherings in a \emph{daily network} \cite{palla2007quantifying, mucha2010community, gauvin2014detecting, rosvall2010mapping, de2015identifying, salnikov2016using}, but here we have chosen to use a simple matching of graph components to underline that social structures in individual snapshots are so obvious that these sophisticated methods are unnecessary.
None of the methods cited above, however, are designed for detecting communities at multiple temporal scales; therefore they are unable to identify cores and help us understand their patterns over time.

Using an individual's cores as a set of social states, we can define a social trajectory which simplifies the description of their social activity, allowing for an exploration of the predictability encoded in social routine. 
We find that, for individuals in our population, times of exploration in the geospatial domain are connected to a small subset of social circles, suggesting a general hypothesis that when humans are most unpredictable with respect to location we are most predictable with respect to our social context.

In summary, our work provides a first quantitative look at the rich long-term patterns encoded in the micro-dynamics of a large system of interacting individuals, characterized by a high degree of order and predictability.
The work presented here provides a new framework for describing human behavior and hints at the promise of our approach. 
While we have focused on predictability, we expect that our work will support better modeling of a multitude of processes in social systems, from epidemiology and social contagion to urban planning, organizational research, and public health.

\section*{Materials}

\subsection*{The dataset}
The proximity interaction network of participants from the Copenhagen Networks Study based on Bluetooth scans, collected using smartphones.
Physical proximity measured via Bluetooth corresponds to distance between 0 and approximately 10 meters, depending on environmental conditions.
As false positives (reported observations of users not actually present) in Bluetooth scans are unlikely, we use a symmetrized (undirected) network of interactions.

\subsection*{Informed consent}
Data collection was approved by the Danish Data Protection Agency, and informed consent has been obtained for all study all participants.

\subsection*{Definitions of entropy}
For an individual $i$ given a sequence of states we define entropy, or uncertainty, in two ways, (1) uncorrelated entropy $S_i^{\text{unc}}=-\sum_j^{N_i} p_j \log_2 p_j$, where $p_j$ is the probability of observing state $j$, captures the uncertainty of the behavioural history without taking the order of visits into account; (2) temporal (or time-correlated) entropy $S_i^{\text{temp}}=-\sum_{T_i' \subset T_i } p(T_i') \log_2 [p(T_i')]$, where $p(T_i')$ is the probability of finding a subsequence $T_i'$ in the trajectory $T_i$, takes both frequency and order of states into account.
From the entropy one can estimate the upper bound of predictability by applying a limiting case of Fano's inequality~\cite{song2010limits,fano1968transmission}: $S_i=H(\Pi_i)+(1-\Pi_i)\log_2(N-1)$, where $H(\Pi_i)=-\Pi_i \log_2(\Pi_i)-(1-\Pi_i)\log_2(1-\Pi_i)$ and $N$ is the number of states observed by person $i$.

\section*{Acknowledgments}
We thank L.\,K. Hansen, P.\,Sapiezynski, A.\,Cuttone, D.\,Wind, J.\,E.\,Larsen, B.\,S.\,Jensen, D.\,D.\,Lassen, M.\,A.\,Pedersen, A.\,Blok, T.\,B.\,Jor\-gensen, and  Y.\,Y.\,Ahn for invaluable discussions and comments on the manuscript and R.\,Gatej for technical assistance.
This work was supported by the Villum Foundation (High Resolution Networks, awarded to S.L.) the UCPH-2016 grant `Social Fabric'.

\renewcommand{\thetable}{S\arabic{table}}
\renewcommand{\thefigure}{S\arabic{figure}}
\renewcommand{\theequation}{S\arabic{equation}}
\renewcommand{\thesection}{S\arabic{section}}

\setcounter{figure}{0}
\setcounter{equation}{0}

\newpage
\thispagestyle{empty}
\phantom{abc} \vspace{8cm}
\centerline{\vspace{0.5cm}\Huge Fundamental Structures of Dynamic Social Networks}
\centerline{\Huge \emph{Supplementary Information}}
\normalsize
\newpage
\section{Summary of main results}

\begin{figure}[!htbp]
\centering
\includegraphics[width=\linewidth]{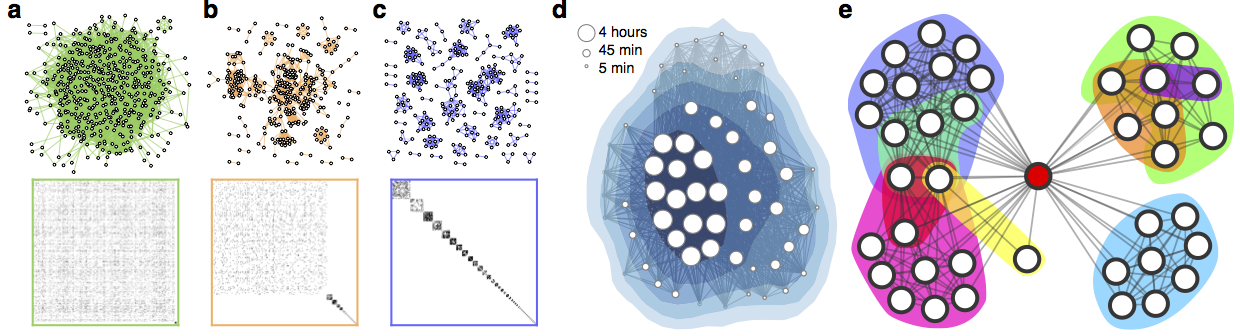}
\caption[Summary of main findings]{\small Summary of main findings. \textbf{a-c}, Network slices obtained by slicing the social dynamics using varying temporal windows (1 day (green), 1 hour (orange), and 5 minutes (blue)). Below, adjacency matrices colored in agreement with networks, and sorted according to component size. \textbf{d}, Gatherings have soft boundaries. The size of each node represents the level of participation. \textbf{e}, Cores simplify social dynamics and provide a context for social interactions.\label{fig:summary}} 
\end{figure}

Figure \ref{fig:summary} illustrates our main findings.
Social groups display a complex temporal behavior, with dynamics spanning multiple time-scales.
Typically, incorporating the temporal dimension drastically complicates the mathematical description of complex networks such that community detection methods require sophisticated mathematical heuristics to disentangle the web of interactions.
By observing social interactions at the right time scale---when the temporal granularity is higher than the turnover rate---we can directly observe social gatherings.
Figure\,\ref{fig:summary}a-c shows social networks obtained using three temporal-windows of increasing size. 
While daily and hourly windows of aggregation obscure social relations (Fig.\,\ref{fig:summary}a-b), a micro-level description directly reveals the fundamental structures.
Applying a simple mathematical matching scheme across time-slices reveals dynamically evolving gatherings with soft boundaries and stable cores (Fig.\,\ref{fig:summary}d).
Unlike the typical community detection assumption of binary assignment, it is clear that some members participate for the total duration of the gathering, while others only participate briefly (Fig.\,\ref{fig:summary}d).
Matching cores across longer time-scales allows us to observe dynamics that unfold over weeks and months.
Cores provide a strong simplification of the social dynamics (Fig.\,\ref{fig:summary}e), and are manifested throughout other data channels such as coordination behavior via call and text messages.
To demonstrate the saliency of our description we use the social contexts provided by cores to quantify the predictability of social life and give a proof of concept of a new type of non-routine prediction.

\subsubsection*{A note on correlations induced by temporality}
\begin{figure}[htb]
	\centering
	\includegraphics[width=\hsize]{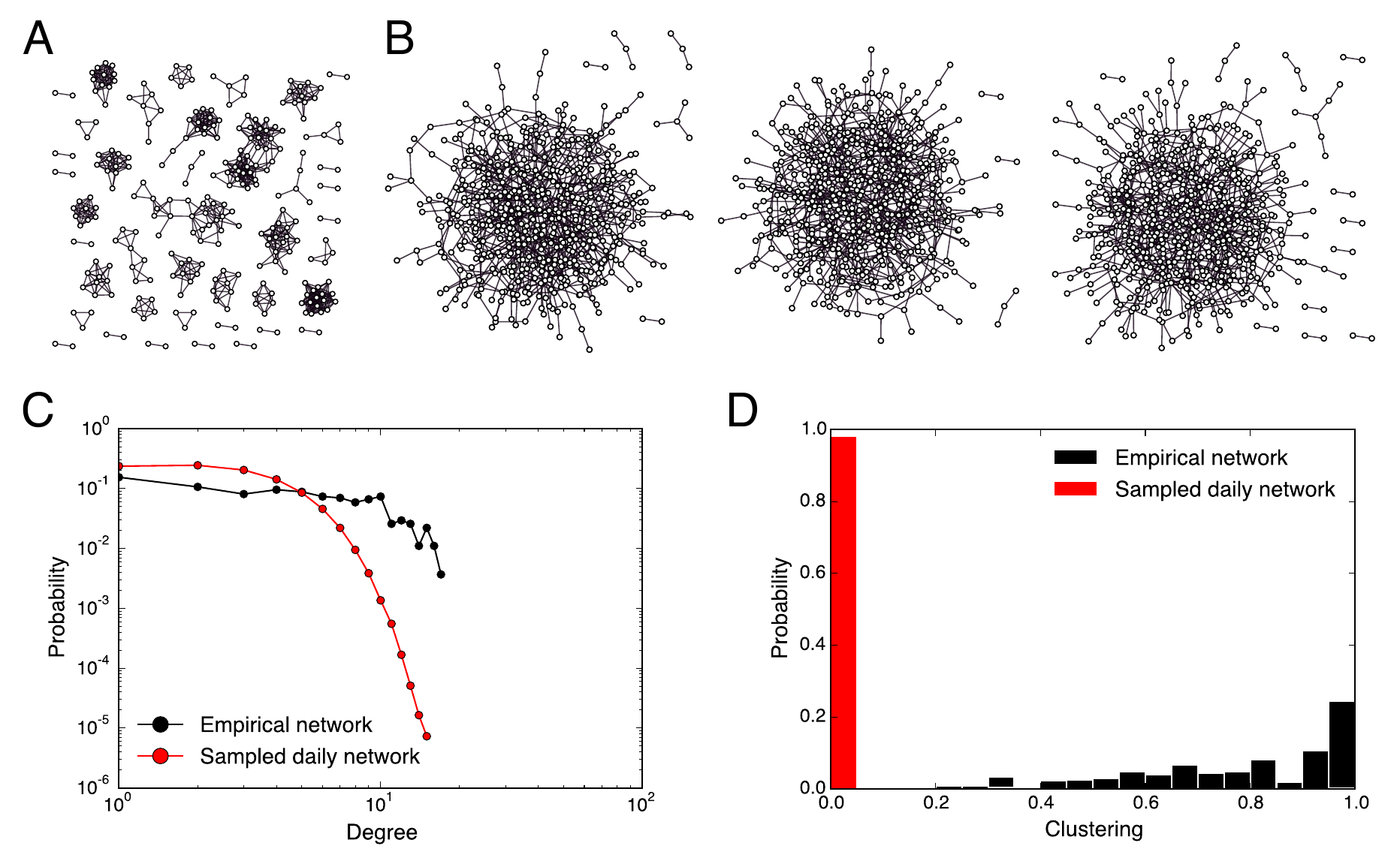}
	\caption[Fundamental difference induced by temporality]{
Fundamental difference induced by temporality. \textbf{A}. A snapshot of social interactions within a 5-minute temporal bin, with $E$ edges.
\textbf{B}. Three reference networks constructed by randomly sampling $E$ edges from the daily interaction graph (Fig 1A, main manuscript) . 
\textbf{C}. Degree distribution for the network in A and 1000 random references networks constructed as described in B. 
\textbf{D}. Clustering distribution. Sampling $E$ edges from the daily network produces mainly one big component with low probability of observing high degree nodes and with low clustering as compared to the temporal snapshot.\label{fig:sampling_hairball}}
\end{figure}

It is instructive to consider the magnitude of the correlations induced by temporality.
We investigate this question as follows.
Let us say that the network in Fig.~\ref{fig:summary}\textbf{c} has $E_c$ edges. 
We now generate a random network based on Fig.~\ref{fig:summary}\textbf{a}, where $E_c$ edges are sampled at random. 
Now we compare the original time-slice (Fig.~\ref{fig:summary}\textbf{c}) with this reference network. 
Fig.~\ref{fig:sampling_hairball} summarizes the key differences. 
Figure~\ref{fig:sampling_hairball}\textbf{A} and~\textbf{B} provide a visual comparison of the networks, which strikingly illustrates how far from a random sample of the aggregated network each temporal network slice is. 
Subfigure \textbf{A} shows the original network from Figure~S1c with $E_c$ edges, and Fig.~\ref{fig:sampling_hairball} shows three examples of subsamples of $E_c$ edges sampled from the network in Figure~S1a. 
We see that whereas the original time-slice has many medium sized components with high clustering, the resampled networks form a single, sparse component with low clustering and a few isolated dyads. 
The basic network statistics confirm how radical this change is. Figure~\ref{fig:sampling_hairball}\textbf{C} shows the change in degree distribution, and Figure~\ref{fig:sampling_hairball}\textbf{D} quantifies the change in the clustering-distribution as we go from the highly clustered real-world data to the tree-like random subsamples. 

\begin{center}
	--- $\bullet$ ---
\end{center}

The remainder of the SI document is organized as follows. Section 2 describes the dataset, Section 3 explains the details of how gatherings are constructed, and describes their basic statistics, including a discussion of dyads. Section 4 shows how cores are extracted and goes into detail on their temporal behavior, as well as the sub-core structures. In Section 5 we provide full details regarding the routine-based geospatial prediction as well as the social context prediction presented in the main text. Section 6 provides background on our model for purely social prediction and, finally, Section 7 discusses background information regarding the coordination leading up to meetings.
\section{Data}
We consider a dataset from the \emph{Copenhagen Networks Study}.
It spans multiple years and measures with high resolution: physical interactions, telecommunications, online social networks, and geographical location. In addition, the dataset contains background information on all participants (personality, demographics, health, politics).  These data are collected for a densely connected population of approximately $1\,000$ students at a large European university.
Data is collected by running custom built applications installed on $1\,000$ smartphones (Google Nexus 4). 
Full details can be found in Ref.\,\cite{stopczynski2014measuring}.

In this manuscript we focus on detecting and tracking co-located groups of individuals during a representative period of five months (roughly one semester), collected between January 1st and June 1st of 2014.
The Bluetooth sensor collects proximity data ($\sim 0-10$ m) of the form $(i,j,t,s)$, where each interaction implies that person $j$ has been in proximity of person $i$ at time $t$, where the devices observe each other with signal strength $s$ \cite{sekara2014strength}.
Bluetooth scans do not constitute a perfect proxy for face-to-face interactions. 
In fact multiple scenarios exist where people in close proximity do not interact and vice versa, nevertheless Bluetooth can successfully be applied in order to sense social networks \cite{stopczynski2014measuring,eagle2009inferring,sekara2014strength}.
Further, our gathering/core-description naturally filters our spurious connections by considering social structures that occur over across extended periods of time.
Gatherings and cores are identified in the proximity network, and the remaining communication channels are used for validation purposes.
In addition, we reserve proximity data from the month of May for validation purposes.
Table \ref{tab:data_stats} shows statistics across the various data sources for 814 individuals, on whom we focus due to their high data quality.

\begin{table}[!htbp]
\centering
\begin{tabular}{l r r}
\toprule
Data source & Total & Unique \\
\midrule
Bluetooth interactions & $14\,673\,869$ & $154\,818$ \\
Call \& text interactions & $75\,364$ & $1\,216$ \\
Geographic locations & $18\,603\,072$ & - \\
WiFi access points & $1\,663\,483\,977$ & $2\,412\,702$ \\
\bottomrule
\end{tabular}
\caption[Data overview]{\small Data overview from January 1st -- June 1st. Bluetooth and call \& text logs are summarized for within-participant relations and do not include external interactions. The unique field denotes the number of distinct observation of each quantity, e.g. number of uniquely observed links.}
\label{tab:data_stats}
\end{table}


\subsection{Construction of temporal network slices}
The data collection application triggers Bluetooth scans every five minutes from the time a phone is turned on; for this reason, the collection of sensors does not follow a global schedule.
To account for this behavior we divide all temporal information into absolute time-windows, $\Delta$ minutes wide.
Within each temporal bin, we draw a unweighted undirected link between two individuals if either $i$ has seen $j$ or vise versa.

\subsection{Selection of time-scales}
\begin{figure}[!htbp]
\centering
\includegraphics[width=0.8\linewidth]{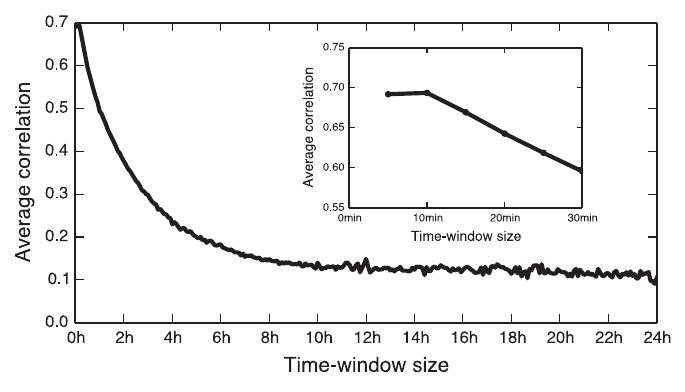}
\caption[Correlation between network slices]{\small Average correlation between network slices, averaged across all time-bins. Inset shows a closeup for the smallest bin-sizes. Calculated for proximity data form March 2014.} 
\label{fig:correlation_slices}
\end{figure}

The scanning frequency of the application sets a natural lower limit of the network resolution to 5 minutes, however, there is no such upper limit for aggregation.
So-called natural timescales have previously been investigated for specific networks with respect to their global topological properties~\cite{clauset2012persistence, sulo2010meaningful, krings2012effects}.
Here we consider the correlation between slices (or turnover of nodes between slices) defined as $C=|E_i^{\Delta} \cap E_{i+1}^{\Delta}|/|E_i^{\Delta} \cup E_{i+1}^{\Delta}|$, where $E_i^{\Delta}$ denotes the set of edges that are observed in bin $i$ with given temporal width $\Delta$.
According to Fig.~\ref{fig:correlation_slices} the average correlation decreases sharply as function of window-size, achieving maximum correlation for small bins.
Slicing network dynamics into short slices (high resolution) also disentangles the network (Fig.~\ref{fig:summary}), thus when time slices are shorter than the group's turnover rate, we can directly observe individuals' group affiliations.
Based on Fig.~\ref{fig:correlation_slices} we chose a temporal width of $5$ minutes, but windows of $10$ minutes could have been chosen without deterioration of results.

When we consider the duration of time-windows, there are two aspects to consider, a \emph{methodological perspective} and a \emph{practical perspective}.
From the methodological perspective, using any finite time window shorter than 10 minutes (as argued in Fig.~\ref{fig:correlation_slices}), will produce analyses and results that are consistent with what we have presented in the main text. 
Using a resolution finer than 5 minutes could potentially be interesting if one was interested in measuring the precise dynamics of group formation (answering questions such as: what is the precise sequence of arrivals?, does a certain group member consistently arrive before everyone else? etc.) or group dissolution (e.g. someone leaving 30 seconds before the rest of the group could be an important social signal). 

Considering the practical perspective, it is clear that measuring e.g.~once every millisecond would produce $300\,000$ times more data than we are currently collecting, while adding very little extra information about the social system. 
For the type of dynamics considered in the main text (e.g.~individuals moving from one gathering to another) 5 or 10 minute time-bins arguably present an useful trade-off between measurement accuracy and the meaningful network changes.

\section{Gatherings}

In this section we describe how connected components in the proximity network are matched across short timescales into dynamical ensembles, which we denote as gatherings.
We then present fundamental statistics on gatherings (size distribution, durations, stability, start/end times), and analyze these properties in the light of on/off-campus behavior. Finally, we focus on identifying repeated gatherings across longer timescales to infer dynamical communities. 

\subsection{Detecting gatherings} \label{sec:detecting_gath}
In each temporal slice we identify connected components, i.e.~nodes that are in close physical proximity as social groups.
Since dyadic relationships qualitatively differ from group relations~\cite{simmel1950quantitative,coser1971masters} we generally treat components of size two separately.

\begin{figure}[!htbp]
\centering
\includegraphics[width=\linewidth]{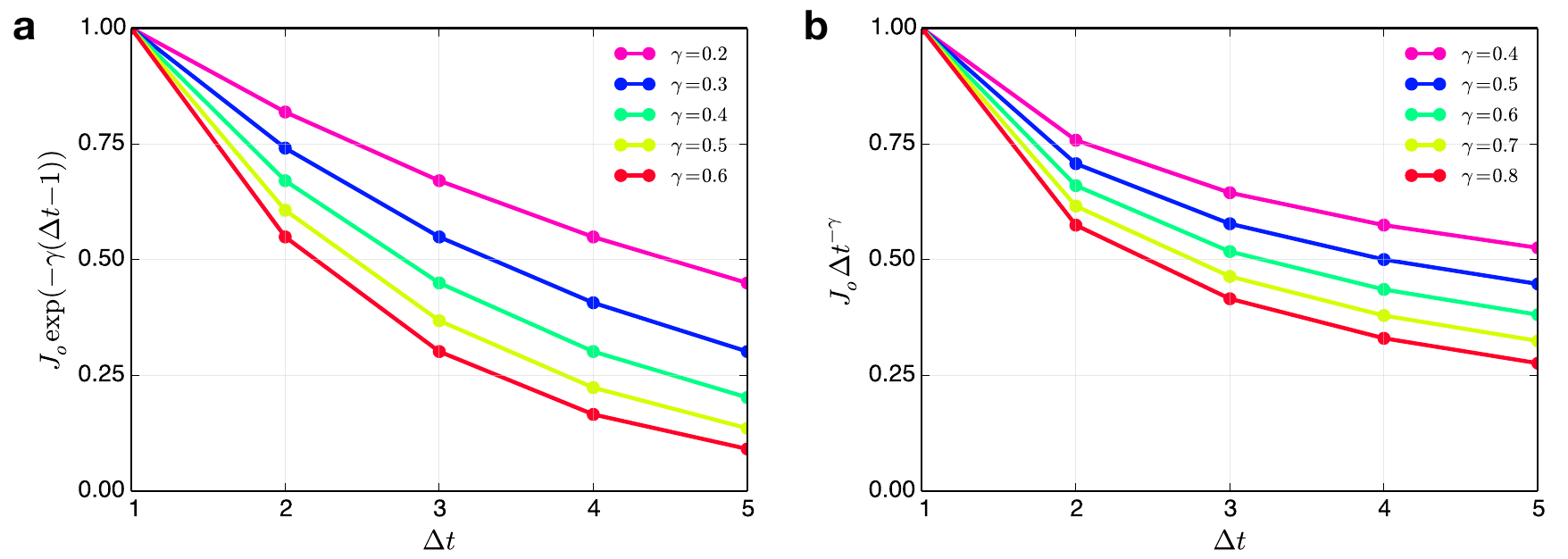}
\caption[Temporal coupling]{\small Temporal coupling between two temporal slices as function of the decay parameter $\gamma$. By definition the decay is zero between consecutive slices ($\Delta t=1$). \textbf{a}, Exponential decay. \textbf{b}, Power law decay.}
\label{fig:decay}
\end{figure}

A gathering is defined as a group that is persistent across time.
To identify gatherings we apply agglomerative hierarchical matching, a widely used method, that merges groups based on their mutual distance (defined below) \cite{ward1963hierarchical, gower1969minimum}, using a matching strategy similar to Green \emph{et al.}~\cite{greene2010tracking}. 
Each group is initially assigned to its own cluster, then every iteration-step merges the two clusters with smallest distance according to the single linkage criteria ($\min(d_i(c_t,c_{t'}))$).
This merge criterion is chosen because it is strictly local and will agglomerate clusters into chains, a preferable effect when clustering groups across time.
The clustering process is repeated until all groups have been merged into a single cluster.
Distance between groups is calculated using a modified version of the Jaccard similarity:
\begin{equation} \label{eq:decay_jaccard}
d(c_t,c_{t'})=1-\frac{|c_{t}\cap c_{t'}|}{|c_{t} \cup c_{t'}|} f(\Delta t,\gamma),
\end{equation}
where $f(\Delta t,\gamma)$ is a term that denotes the coupling between temporal slices and $\Delta t = t' - t$ denotes the temporal distance between two bins (for consecutive bins $\Delta t =1$).
The function can assume any form, increasing or decreasing; we utilize it to model decay between temporal slices, with the two most prominent forms being: exponential ($\exp \left(-\gamma \left( \Delta t-1 \right) \right)$), and power-law ($\Delta t^{-\gamma}$), see Fig.~\ref{fig:decay}.
Thus, by definition the term assumes the value 1 (zero decay) between two consecutive temporal slices.
For computational reasons we only focus on gatherings identified using exponential decay with $\gamma=0.4$, other decay parameters yield similar results see SI Section \ref{sec:temp_decay}.

\subsubsection{Partitioning the dendrogram} \label{sec:partition_denrogram}
\begin{figure}[!htbp]
\centering
\includegraphics[width=\linewidth]{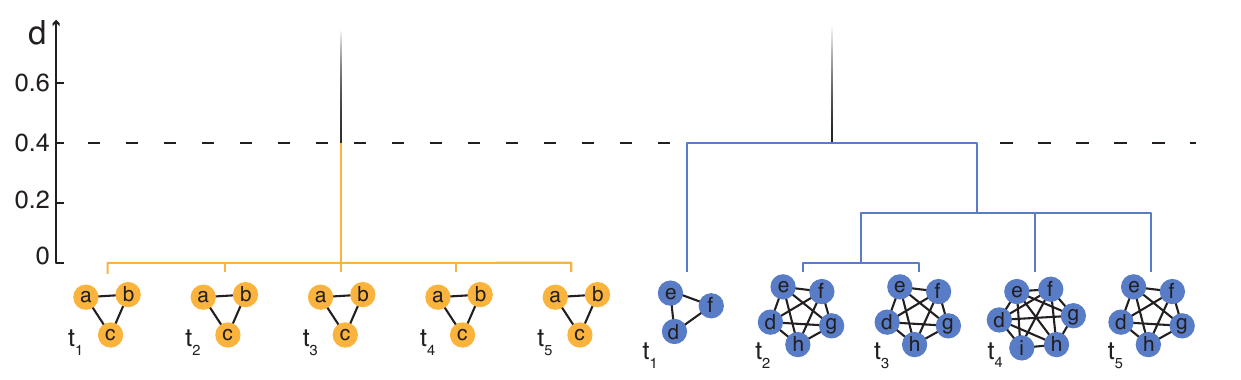}
\caption[Illustration of group dendrogram]{\small Illustration of constructed dendrogram, depicting distance ($d$) between groups identified across 5 timebins. The tree is constructed using an exponential decay function with $\gamma=0.4$. Two gatherings, orange and blue, are inferred by thresholding the tree, where all groups below or equal to the threshold ($d=0.4$) are merged.}
\label{fig:dendrogram}
\end{figure}

\begin{SCfigure}[][!htbp]
\centering
\caption[Stability measures of gatherings]{\small Stability measures of gatherings. \textbf{a}, Illustration of the local stability measure, calculated between consecutive slices. \textbf{b}, Global measure calculated between each slice and the aggregated structure. \textbf{c},  Global ($\sigma$) and local ($\eta$) stationarity of gatherings as a function of partition value $d$, averaged over all gatherings identified in January 2014. Gatherings achieve optimal stability around $d\sim 1/2$.}
\includegraphics[width=0.7\linewidth]{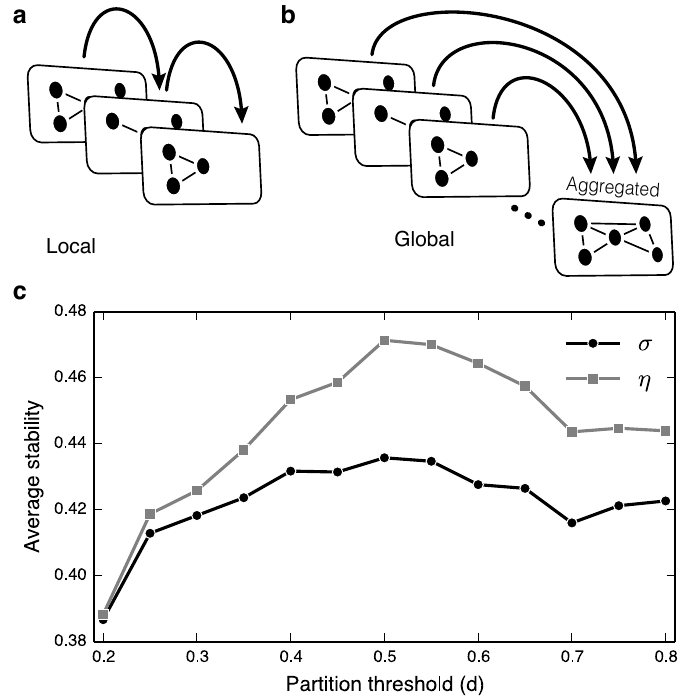}
\label{fig:stability_measures}
\end{SCfigure}

The method described above iteratively constructs a dendrogram where temporally localized groups are hierarchically clustered.
To extract meaningful social structures we need to partition the dendrogram (Fig.~\ref{fig:dendrogram}).
Modularity and partition density have previously been applied for similar purposes \cite{newman2004finding,ahn2010link}, but these do not generalize well for temporal processes.
Instead we consider the cluster stability with respect to local and global measures.
Local stability ($\eta$) is calculated from the average node-wise overlap between consecutive slices (Fig. \ref{fig:stability_measures}a), while global ($\sigma$) is calculated from the average overlap between all slices and the aggregated structure (Fig. \ref{fig:stability_measures}b):
\begin{eqnarray} \label{eq:stability1}
\eta&=&\frac{\sum\limits_{t=t_{\text{birth}}}^{t_{\text{death}}-1} J(g_t,g_{t+1})}{t_{\text{death}}-t_{\text{birth}}-1}, \\
\label{eq:stability2}
\sigma&=&\frac{\sum\limits_{t=t_{\text{birth}}}^{t_{\text{death}}} J(g_t,G)}{t_{\text{death}}-t_{\text{birth}}},
\end{eqnarray}
where $t_{\text{birth}}$ and $t_{\text{death}}$ are respectively the birth and death of the gathering, $g_t$ is a temporal slice, $G=g_{\text{birth}} \cup g_{\text{birth}+1} \cup \cdots \cup g_{\text{death}}$ is the aggregated structure, and $J$ is the overlap ($J=|i\cap j| / |i \cup j|$), defined as zero if the gathering has only existed for one time bin.
Palla \emph{et al.}~\cite{palla2007quantifying} applied a related measure to estimate the stationarity of communities.
Varying the partition threshold (Fig.~\ref{fig:stability_measures}c) we observe a maximum in both measures, indicating a regime where gatherings are both temporally and globally stable.
Threshold values of $d \geq 1/2$ are, however, problematic since they merge gatherings that split into two equally sized parts together with both parts, or vice versa.
In this scenario, we find that the desirable behavior is to declare the old gathering as `dead' and identify two new gatherings as born. 
Therefore, to achieve optimal stability and to avoid issues with unwanted merging we partition the dendrogram at $d=0.49$.

\subsubsection{Temporal decay function} \label{sec:temp_decay}
\begin{SCfigure}[][!htbp]
\centering
\includegraphics[width=0.7\linewidth]{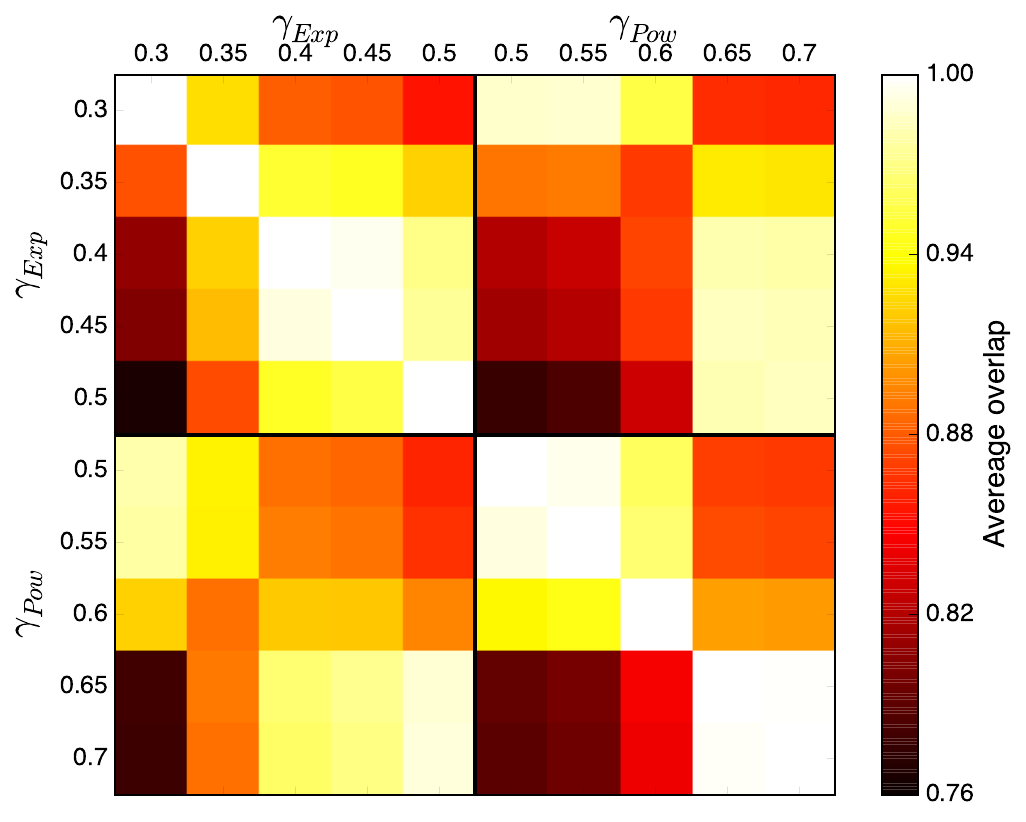}
\caption[Effect of decay parameter on robustness ]{\small Effect of decay parameter on gathering robustness, calculated using Equation \ref{eq:overlap}. Colorbar shows average overlap between two sets of gatherings, and never drops below $0.76$, indicating the robustness of the procedure with respect to parameter perturbations.}
\label{fig:gathering_stability}
\end{SCfigure}

Here we investigate how robust the inferred gatherings are to perturbation of the $\gamma$-parameter and using an alternate decay form. 
To compare two set of gatherings (identified using different $\gamma$-values) we calculate the average maximal overlap between individual gatherings as
\begin{equation} \label{eq:overlap}
O_{\gamma\gamma'}=\frac{1}{|G_{\gamma}|}\sum\limits_{i\in G_{\gamma}} \max_{j \in G_{\gamma'}} (J(i,j)),
\end{equation}
where $G_{\gamma}$ denotes the set of gatherings found using a specific value of $\gamma$, and $|G_{\gamma}|$ is the number of gatherings.
Overlap is calculated using Jaccard similarity, $J=|i\cap j|/|i \cup j|$, where $i$ and $j$ are respectively gatherings from $G_{\gamma}$ and $G_{\gamma'}$.
Figure~\ref{fig:gathering_stability} shows the overlap matrix between identified sets of gatherings; in general it assumes overlap values above $0.76$ for any choice of parameters.
Indicating that the gatherings are robust to even large perturbations. 

\subsubsection{Gathering timescales}
The outlined method identifies multiple gatherings, some only exist momentarily while others are sustained for long time periods.
One can easily imagine brief encounters between good friends as being more meaningful than prolonged interactions between individuals commuting to work; this therefore raises the question of which meetings are meaningful and which are not.

Here we simply adopt the convention developed by the \emph{Rochester Interaction Record} \cite{reis1991studying}, where meaningful encounters are defined as those lasting $10$ minutes or longer. In order to filter out spurious connections, we impose the requirement that a gathering must be observed for at least $4$ consecutive time slices to be represented in our statistics.

\begin{figure}[!htbp]
\centering
\includegraphics[width=0.9\linewidth]{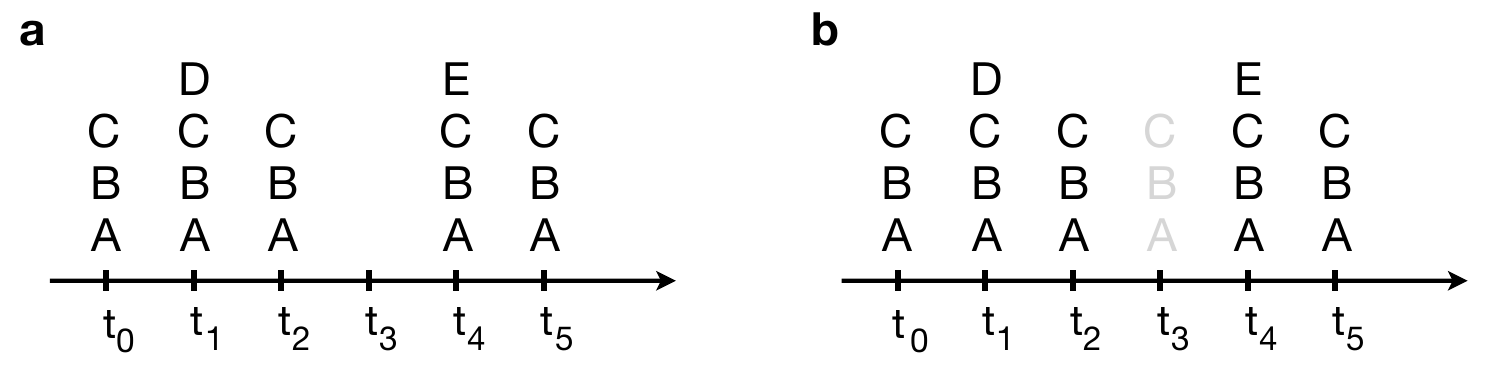}
\caption[Gathering micro fluctuations]{\small Gathering micro fluctuations. \textbf{a}, A gathering appears, disappears, and re-appears moments later. \textbf{b}, If this occurs in rapid succession we treat the gathering as if is was present in bin $t_3$ with the same nodes as in bin $t_2$.}
\label{fig:absent}
\end{figure}

While dynamics on $20$ minutes+ timescales describe the overall evolution of a gathering, micro dynamics on 5-minute scales represent everyday events such as going to the bathroom.
Gatherings, therefore, might disappear and reappear within very short time-intervals (Fig.~\ref{fig:absent}a), in such cases we use imputation, see Fig.~\ref{fig:absent}b.

\subsection{On- \& off-campus gatherings} \label{sec:on_off}
Gatherings are not geographically constrained and therefore free to occur anywhere, in this section we focus on distinguishing between gatherings that occur on- and off-campus.
We do this by applying data acquired through the WiFi channel, where each mobile phone scans for nearby wireless network access points (AP) every 5 minutes or less and logs their unique identifier and name.
The entire university campus is densely covered by wireless networks and it is therefore highly unlikely that students located on campus will not see an university AP.
Because the names of campus APs follow a uncommon and standardized naming scheme we can infer when a student is close to a campus access point and use this as a proxy of being on or off campus.
For each participant we construct a 5-minute binned vector containing binary values 0 (not on campus) and 1 (on campus).
Because gatherings are an ensembles of people we perform majority voting across nodes for each time-bin to determine whether the gathering was on campus or not in that specific time-bin.
Since gatherings are spatio-temporal entities, we also perform a majority voting across all time-bins to achieve a hard split and determine whether a gathering mainly occurred on- or off-campus.
This yields $13\,872$ off- and $9\,195$ on-campus gatherings and with an $92.84\%$ average agreement between votes.
The primary reason for disagreement are mobile gatherings traveling to or from campus (e.g.~on a bus or walking).

\subsection{Gathering statistics}
\begin{figure}[!htbp]
\centering
\includegraphics[width=\linewidth]{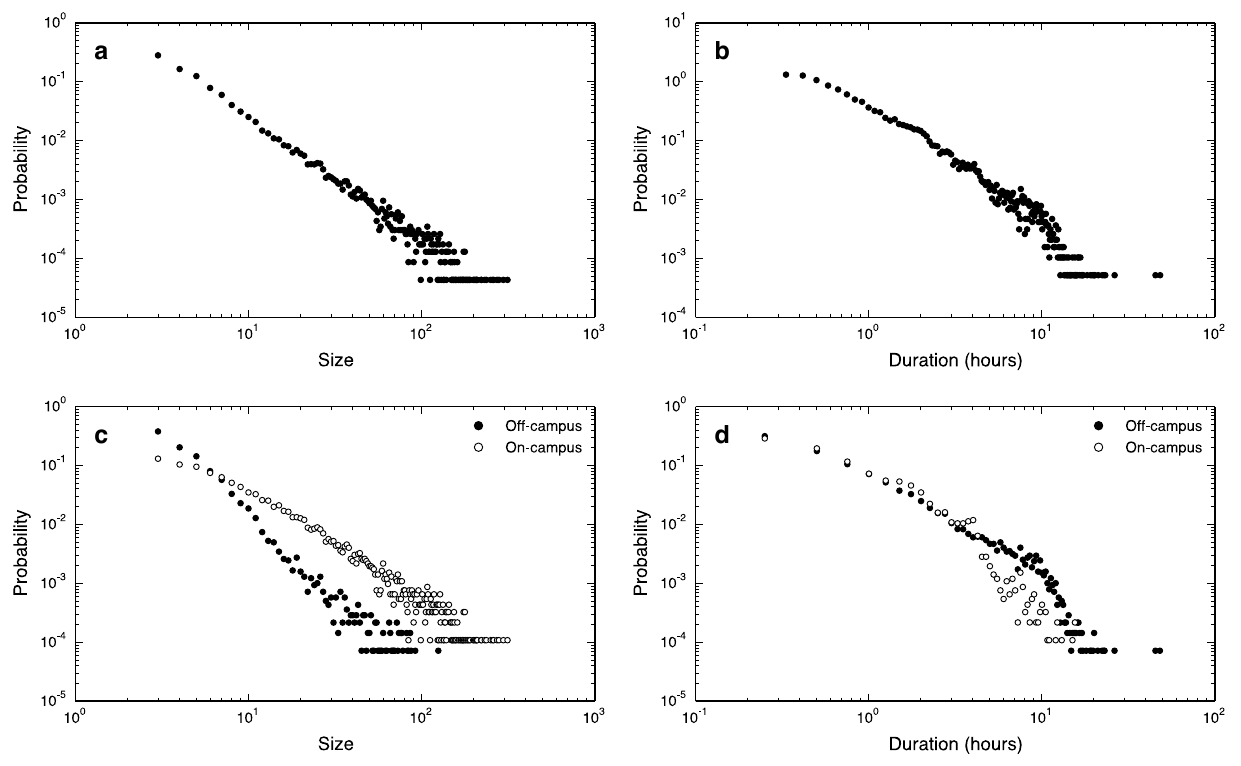}
\caption[Statistical features of gatherings]{\small Statistical features of gatherings. Summarizing size and duration distributions for all gatherings (panels a-b) and conditioned on off/on-campus meetings (c-d).  \textbf{a}, Gathering size distribution. \textbf{b}, Gathering duration distribution. \textbf{c}, Gathering size distribution for on- and off-campus meetings. \textbf{d}, Gathering duration distribution, based on location (on/off-campus).\label{fig:gath_stat}}
\end{figure}

Gatherings show a broad distribution in both size and duration, see Fig.~\ref{fig:gath_stat}.
Dividing gatherings into on- and off-campus categories reveals that meetings occurring on university campus are larger (Fig.~\ref{fig:gath_stat}c), but have considerably lower probability of lasting longer than 4 hours (Fig.~\ref{fig:gath_stat}d). 
This suggests that large meetings are mainly driven by the class schedule, while meeting duration is determined by social context.
Because meetings occurring outside of campus often require increased coordination, our hypothesis is that once groups meet, they will interact across longer periods of time for the meeting to pay off with respect to the organizational cost. 

\begin{figure}[!htbp]
\centering
\includegraphics[width=\linewidth]{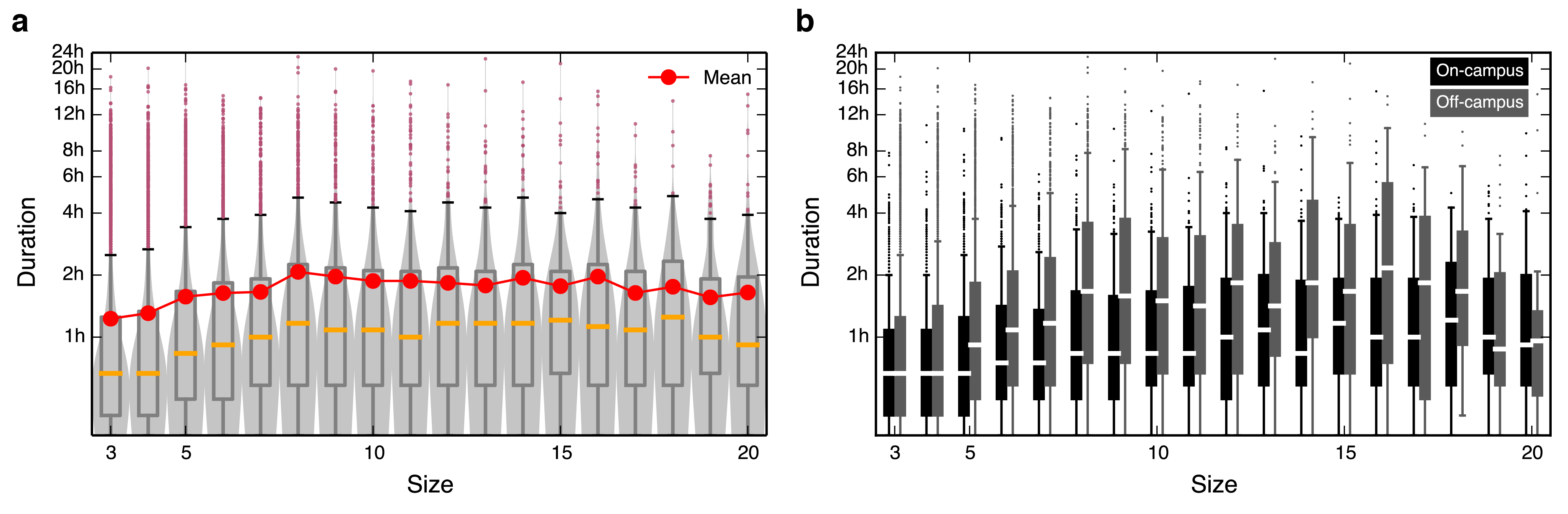}
\caption[Duration as function of gathering size]{\small Duration as function of gathering size. \textbf{a}, Violin plot shows the distribution of durations as function of size, summarized across all gatherings. \textbf{b}, Box plot of the duration distributions divided into on- and off-campus meetings.}
\label{fig:gath_dur_size}
\end{figure}

It is also interesting to consider the duration of each meeting as function of the total number of nodes that participate in it.
Figure \ref{fig:gath_dur_size}a shows broad distributions of duration across all sizes, however, both the mean and median are quite stable and reveal that small gatherings on average have shorter durations compared to larger meetings.
Further dividing the data into on- and off-campus categories (Fig.~\ref{fig:gath_dur_size}b) shows that small meetings on and off campus are quite similar with respect to duration, while larger meetings tend to last longer, provided that they occur off-campus. 

\begin{figure}[!htbp]
\centering
\includegraphics[width=\linewidth]{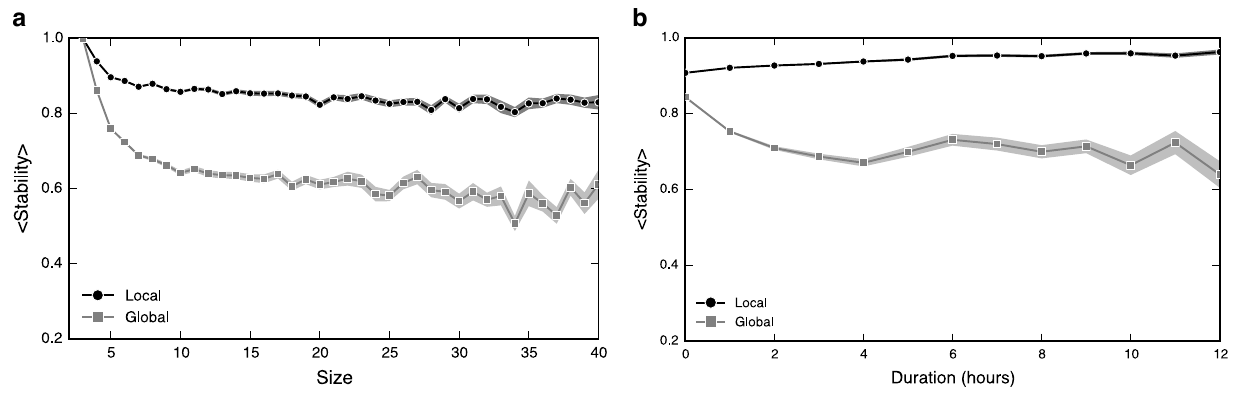}
\caption[Gathering stability versus size and duration]{\small Global and local gathering stability. \textbf{a}, Average stability as function of size, averaged across all gatherings with a specific size. Full lines denotes mean, while shaded areas shows the standard deviation of the mean. \textbf{b}, Average stability as function of duration, binned into one hour wide bins. Fully drawn lines denote the mean, while shaded areas illustrates its deviation.}
\label{fig:gath_stab}
\end{figure}

Combining the statistics above with Eqs.~(\ref{eq:stability1}-\ref{eq:stability2}) we can explore how gathering stability depends on size and duration of meetings.
Figure \ref{fig:gath_stab}a reveals that the local churn between consecutive time slices is quite constant, irrespective of size, indicating that there is a low turnover of nodes between slices---on average much lower than predefined by the partition threshold.
For small gatherings, however, we observe finite size effects due to the partition threshold (see sec. \ref{sec:partition_denrogram}). 
Global stability is lower, but also fairly independent of gathering size.
With respect to duration (Fig.~\ref{fig:gath_stab}b), local stability increases as meetings duration increases, revealing that longer meeting have lower turnover of nodes between consecutive slices.
The global measure shows similar behavior. It achieves a slightly lower stability and shows that gatherings are globally stable independent of duration.
This combination of a constant turnover between slices and a global stability suggests that gatherings contain groups of highly interacting individuals, that are present throughout the entirety of the meeting, while other individuals participate infrequently and are constantly being replaced.
Similar social structures have previously been observed in the social science literature where the individuals have been defined as core members \cite{davis1941deep}. 

\begin{figure} [!htbp]
\centering
\includegraphics[width=\linewidth]{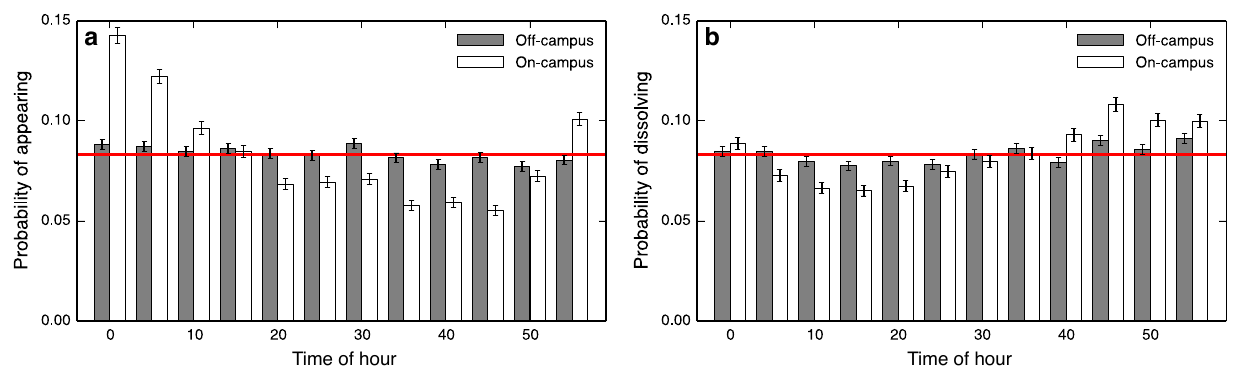}
\caption[Temporal patterns of gatherings]{\small Summary statistics of gathering temporal patterns, summarized for on- and off-campus meetings. Red line denotes the uniform probability distribution for the case where all states are equally probable. \textbf{a,} Probability of a gathering appearing, calculated from the first time-bin in which we observe it. \textbf{b,} Probability of dissolving, i.e. the last time-bin where we observe a gathering.}
\label{fig:gath_timebins}
\end{figure}

Finally we can look into specific temporal patterns with regards to when on and off campus gatherings occur, i.e. in which 5 minute time-bins they first appear and later disappear.
Figure \ref{fig:gath_timebins}a reveals that on-campus gatherings have increased probability of occurring exactly on the hour, while off-campus meetings are evenly distributed---clearly showing the effect of the class schedule. 
A similar case is seen for the probability of dissolving (Fig.~\ref{fig:gath_timebins}b), where on-campus meetings mainly end on integer hour values.  

\subsection{Temporal communities} \label{sec:matching}
\begin{figure}[!htbp]
\centering
\includegraphics[width=\linewidth]{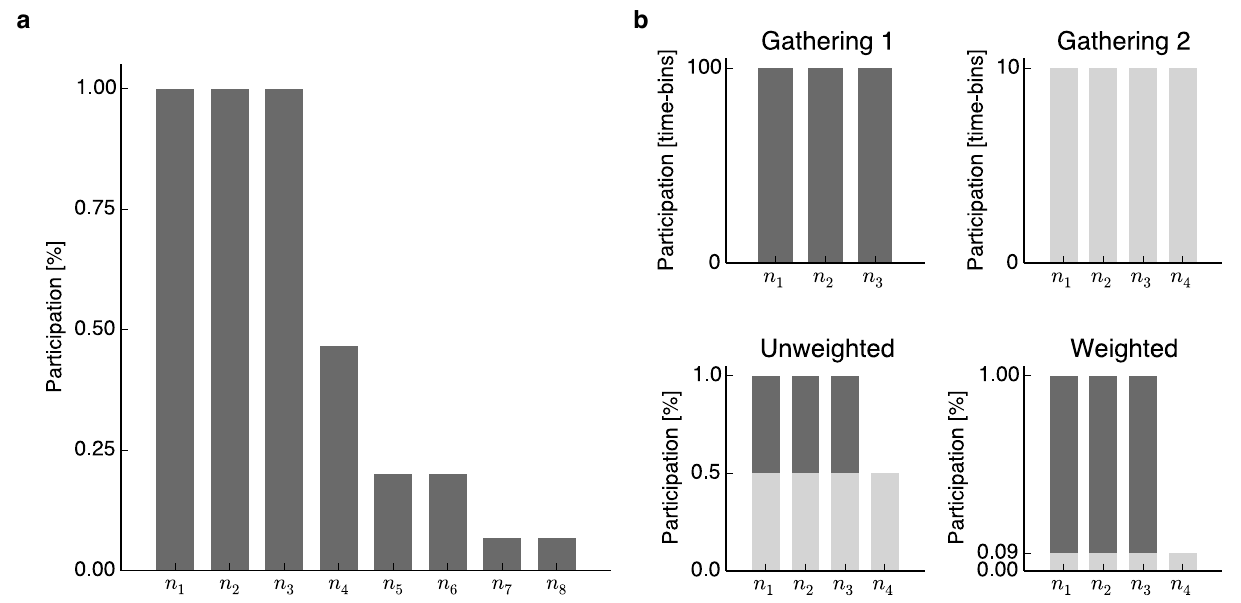}
\caption[Gathering participation profile]{\small Illustration of gathering participation profiles. \textbf{a}, Each vertical bar indicates the overall fraction of time a node has spent in the gathering. Values are normalized according to the total gathering lifetime. \textbf{b}, Illustration of how to construct a community participation profile based on its constituent gatherings. Top figures depict two gatherings with unequal lifetimes. Bottom plots illustrate the principles behind the unweighted and weighted methods. Bar colors indicate the role each node plays in the community profile.}
\label{fig:gathering}
\end{figure}

So far each gathering only contains information about its local appearance, to gain a dynamical picture we match gatherings across time.
Due to soft boundaries, a strict matching criteria is not a feasible method, since a person who coincidently walks past a group might be included in it. 
Thus, we expect noise to be present in each gathering.
To mitigate this effect we instead match gatherings according to the participation levels of their constituent nodes.
Counting the fraction of times a nodes has been present in the gathering we construct a normalized participation profile, see Fig.~\ref{fig:gathering}a.
Again, gatherings are matched according to their individual participation profiles using agglomerative hierarchical clustering.
Since nodes, however, no longer assume binary values but may assume participation levels in the interval $0 < n_i \leq 1$, we calculate distance based on a continuous version of the Jaccard similarity:
\begin{equation} \label{eq:cont_jaccard}
D\left( G_i,G_j \right) = 1 - \frac{\sum\limits_{n=1}^N \min \left( G_i ,G_j \right)}{\sum\limits_{n=1}^N \max \left( G_i , G_j \right)},
\end{equation}
where $G_i$ is a vector containing node-wise participation values for gathering $i$, and $N$ is the total number of nodes in $G_i \cup G_j$.
The two functions $\max$ and $\min$ act piecewise on the two vectors, and $D(G_i,G_i)$ is defined as $1$ between two gatherings that have zero overlap. 
When merging clusters of gatherings $(\mathcal{G})$ we apply the average linkage criterion and define the average distance between them as
\begin{equation}
D(\mathcal{G},\mathcal{G}')=\frac{1}{ | \mathcal{G} | |\mathcal{G}'| }\sum\limits_{G\in \mathcal{G}} \sum\limits_{G'\in \mathcal{G}'} D ( G,G' ),
\end{equation}
where $|\mathcal{G}|$ denotes the cardinality of a set of gatherings.
Other linkage criteria can also be used, such as complete or Ward-linkage. 
Iteratively this method builds a dendrogram with gatherings as leafs.
Thresholding the tree partitions similar gatherings together into communities.
A community then consists of all nodes from its constituent gatherings, but it also inherits their individual participation profiles.
Thus we need a method to construct a community participation profile from its subcomponents.
This can be done using two methods: weighted or unweighted, with the differences illustrated in Fig.~\ref{fig:gathering}b.
The unweighted method takes into account the gathering participation profiles and calculates the average, weighing each gathering equally:
\begin{equation}
C=\frac{1}{|\mathcal{G}|}\sum\limits_{G\in\mathcal{G}} G.
\end{equation}
The weighted version instead assigns each gathering a weight according to its lifetime ($\tau_{\text{life}}$), i.e. number of temporal bins it has been present:
\begin{equation}
C^{\text{weighted}}=\frac{1}{\sum \tau_{\text{life,G}}}\sum\limits_{G\in\mathcal{G}} \tau_{\text{life,G}} G.
\end{equation}
Both measures comparatively construct similar dynamical communities and yield similar overall statistics; we choose the the weighted version, because it is slightly less influenced by noise.

\subsubsection{Optimal clustering partition}
\begin{figure}[!htbp]
\centering
\includegraphics{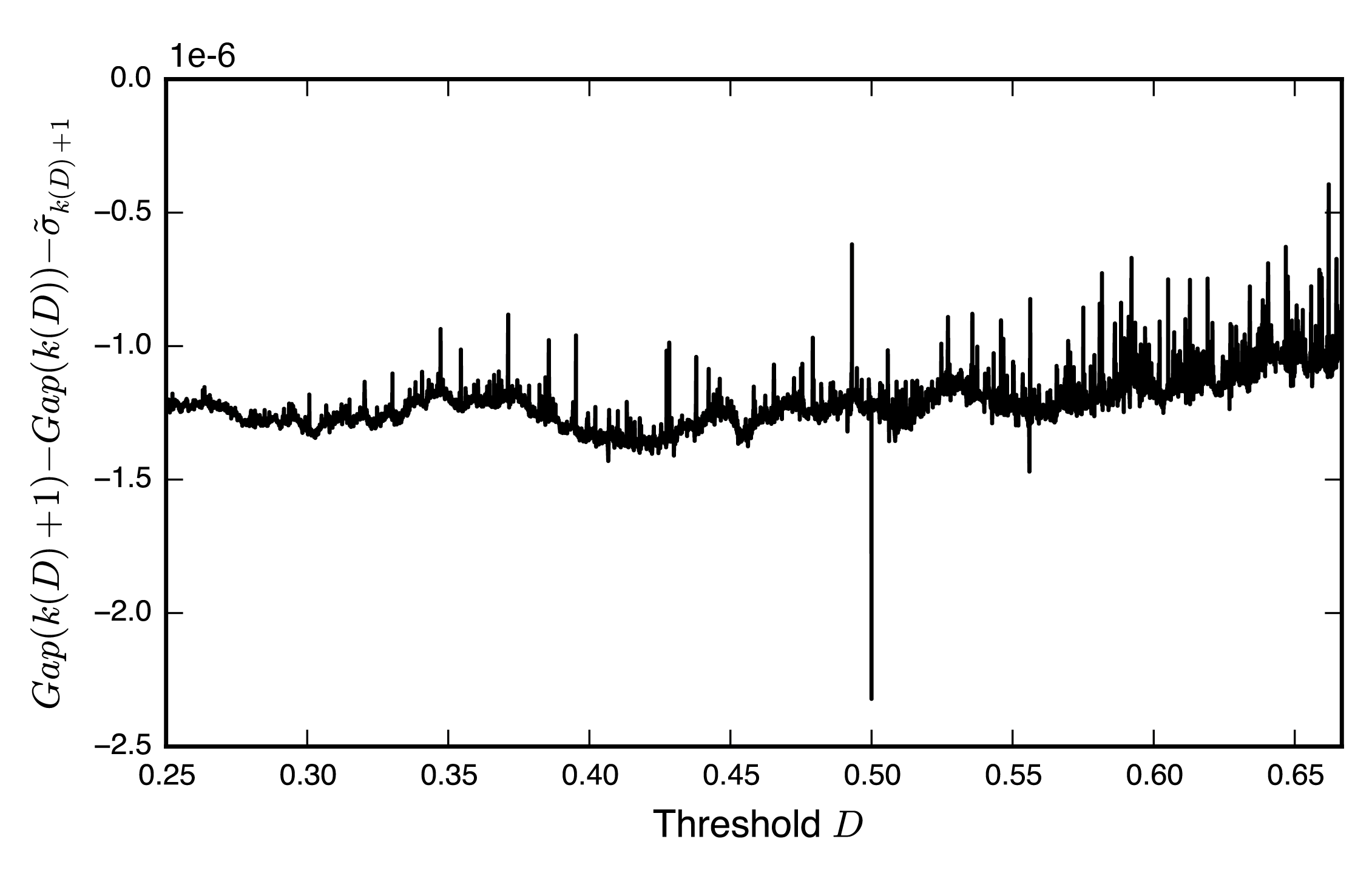}
\caption[Estimating optimal number of clusters using the Gap measure]{\small Estimating the optimal number of clusters using the gap measure, with the number of clusters $k$ being directly related to the threshold value $D$.}
\label{fig:gap}
\end{figure}

Applying hierarchical clustering to merge similar gatherings into communities again leaves one open question: Which partition value is the optimal?
Using a similar line of reasoning to what we presented in in Sec. \ref{sec:partition_denrogram}, we can argue that a threshold value of $D=0.49$, is preferable.
It is, however, possible to estimate the exact optimal threshold value, which we can compare to our hypothesized guess. 
Applying a heuristic inspired by the Gap statistic introduced by Tibshirani \emph{et al}.~\cite{tibshirani2001estimating}, we compare the clustering according to a null model distribution (for a comprehensive survey of methods see Milligan and Cooper \cite{milligan1985examination}).
Given a total of $m$ gatherings clustered into $k$ clusters (communities): $C_1, C_2,\ldots, C_k$  we calculate the within-dispersion measure as,
\begin{equation}
W_k=\sum\limits_{r=1}^k\frac{1}{2|C_r|} \sum\limits_{i,j\in C_r} D_{ij},
\end{equation}
where $D_{ij}$ is defined in equation~\ref{eq:cont_jaccard} and denotes the pairwise distance between gatherings $i$ and $j$ that both belong to cluster $C_r$.
Again $|\cdot|$ denotes the cardinality of a cluster, i.e. the number of gatherings that are clustered in $C_r$.
The factor $2$ takes double counting into account.
Thus $W_k$ is the accumulated within cluster sum of differences around the cluster mean. 
Applying the principles from Tibshirani \emph{et al.}~\cite{tibshirani2001estimating} we compare $\log (W_k)$ to an expected value generated by a null model distribution of the data.
The gap measure is defined as
\begin{equation}
\text{Gap}(k)=1/B\sum\limits_{b=1}^B \log(W_{kb})-\log(W_k),
\end{equation}
where $B$ denotes the number reference data sets.
The optimal number of clusters is then the value of $k$ for which $\log(W_k)$ falls furthest below the reference data curve, i.e. the value of $k=k^*$ such that
\begin{equation}
k^*=\argmin_{k} \{k|\text{Gap}(k) \geq \text{Gap}(k+1)-\tilde{s}_{k+1}\},
\end{equation}
where $\tilde{s}_k=s_k\sqrt{(1+1/B)}$ and $s_k$ is the standard deviation of $\log (W_{kb})$ over the $B$ synthetic datasets.
Each null model is constructed by assigning random participation values, chosen from a uniform distribution, to random nodes, thus creating reference gatherings with similar size distributions.
According to Fig.~\ref{fig:gap} the gap statistic achieves a minimum, indicating that the optimal place to cut the dendrogram is at distances of $D=0.50$, in good agreement with the previously stipulated value.
A threshold value of $D=0.50$ is, however, problematic since it will cluster gatherings with $50\%$ overlap.
To avoid this issue we cut our dendrogram at $D=0.49$, merging $23\,067$ gatherings into $7\,320$ distinct dynamic communities.

\subsection{Dyadic relations} \label{sec:dyads}

Dyads are a special case of social relations and are hypothesized to be qualitatively different from groups \cite{simmel1950quantitative,coser1971masters}, but the way we identify them is similar to the outlined framework in Section \ref{sec:detecting_gath}.
Since dyads can only be components of size two, tracking their evolution is trivial, because we can apply a strict merge criterion, i.e.~require $100\%$ overlap.
This also applies in the case when tracking repeated appearances (dyadic cores) across the full duration of the dataset.
Thus, they only distinction between a gathering and a dyadic gathering lies in the fact that size of the gathering will always be fixed, meaning if a dyad evolves into a group (of any size larger than two) then we claim that the dyad in question has ended and a new group relation has been initiated.
In accordance with gatherings, we require dyadic gatherings to be present for at least four consecutive bins ($20$ minutes) to be considered as meaningful.
In total we observe $34 996$ dyadic gatherings and $4 844$ unique dyads.

\subsubsection{Dyad statistics}
\begin{figure}[!htbp]
\centering
\includegraphics[width=\linewidth]{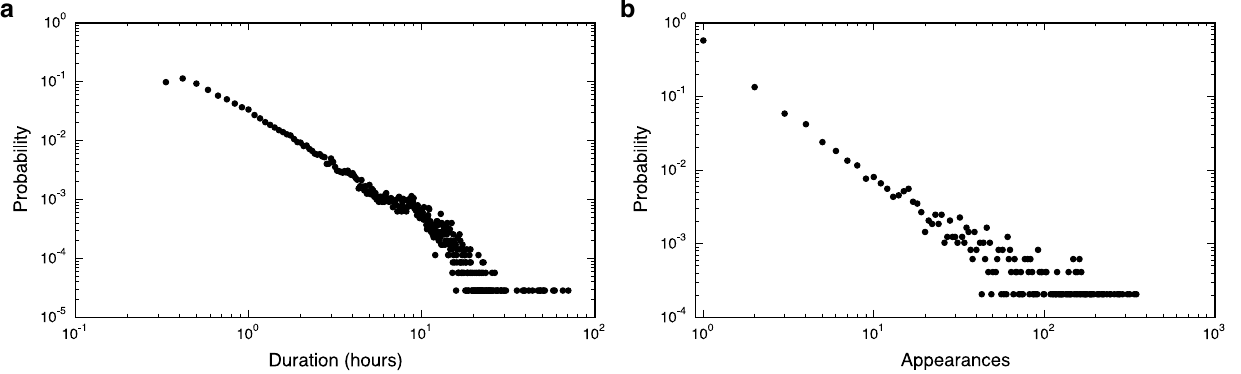}
\caption[Statistical features of dyadic gatherings]{\small Statistical features of dyadic gatherings. \textbf{a}, The distribution of duration for dyadic gatherings. \textbf{b}, The distribution of the number of appearances.}
\label{fig:dyads}
\end{figure}

Similar to group relations, dyadic gatherings produce a broad distribution of durations comparable to the gatherings of groups (Fig.~\ref{fig:dyads}a).
In addition, dyads also produce a broad distribution of repeated appearances, see Fig.~\ref{fig:dyads}b, with a majority of gatherings only being observed once, while others on average can appear multiple times per day.

\section{Cores} \label{sec:cores}
In the previous section (see Fig.~\ref{fig:gath_stab}b) we found that certain individuals are present for a majority of a gathering's lifetime.
In this section, we begin by describing how such \emph{cores} are extracted from temporal communities.
Then, we describe the fundamental statistics of cores (distribution of appearances, distribution of sizes, individual membership in cores, subcore-statistics), as well as the key differences between work- and recreational cores.
Finally, we study the temporal patterns and entropy of core-appearences.

\subsection{Extracting cores}
Nodes within each community have varying attendance (see Fig.~\ref{fig:gathering}a), some are only members for a limited time, while others interact over extended periods of time.
Thus, participation profiles show pronounced core structures, highlighting individuals that act as `generators' of each community.

\begin{figure}[!htbp]
\centering
\includegraphics[width=\linewidth]{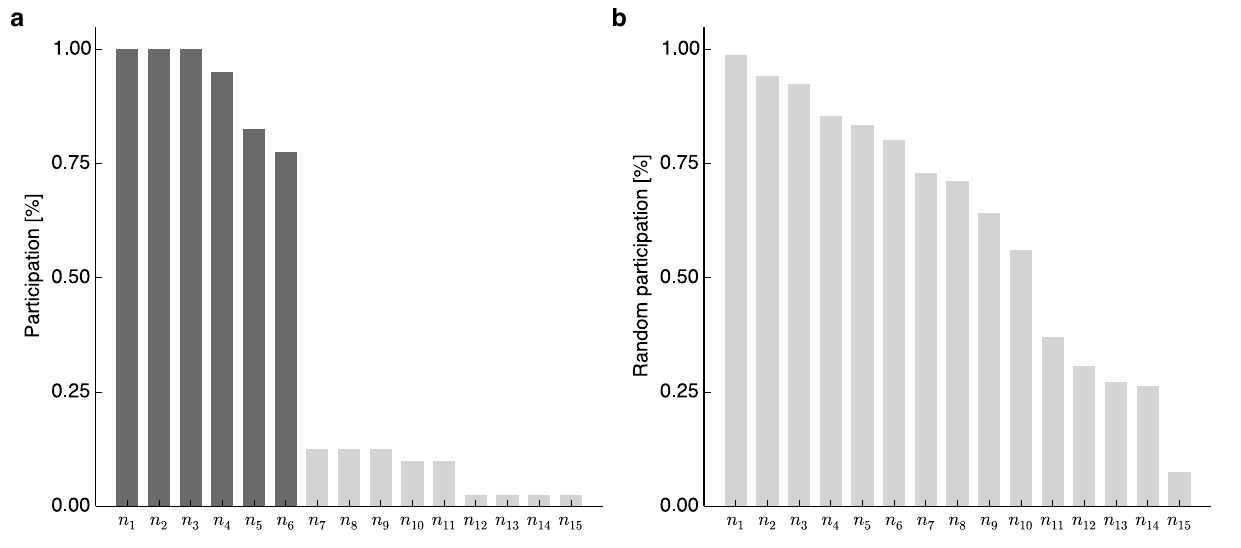}
\caption[Extracting cores]{\small Extracting cores from community participation profiles. Dark gray bars denote nodes with participation levels above the maximal gap. \textbf{a}, Ordered participation profile for a community composed of 15 individuals. \textbf{b}, Similar as in panel a but generated from a uniform random distribution.} 
\label{fig:gap_criteria}
\end{figure}

\begin{SCfigure}[][!htbp]
\centering
\includegraphics[width=0.5\linewidth]{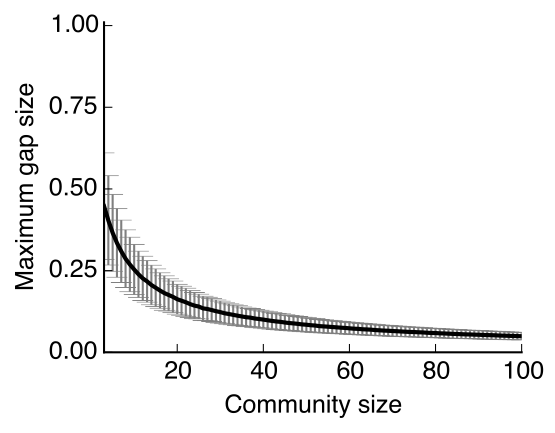}
\caption[Core selection boundary]{\small Core selection boundary. Decision boundary calculated from $N=10\,000$ independent trials for each size. Black line denotes the mean max gap value while error bars indicate the standard deviation.} 
\label{fig:decision_boundary}
\end{SCfigure}

Consider the participation levels of individuals as ordered profiles (Fig.~\ref{fig:gap_criteria}a), where each bar denotes the fraction of time a node has has spent in a community relative to the community's total lifetime.
A significant gap in this profile identifies core nodes.
We compare this to a participation profile generated by a random process (Fig.~\ref{fig:gap_criteria}b), where we pick a random participation level between $0$ and $1$ (for each node) from a uniform distribution.
The maximal gap in this random profile thus tells us whether the real gap is significant.
We can estimate the average expected gap size and deviation by generating many ($N=10\,000$) random participation profiles.  
Generalizing this notion to all sizes of communities we evaluate how significant a gap is compared to the expected value generated at random.
The decision boundary in Fig.~\ref{fig:decision_boundary} divides gap sizes into two regions.
If the actual gap is greater than the average null-model gap $\mu_{\text{random}}$ plus one standard deviation $\sigma_{\text{random}}$, we define the core to be significant. 
Thus, we only keep cores with gap sizes above $\mu_{\text{random}}+\sigma_{\text{random}}$.
According to this criterion, we find that $7\,146$ out of the $7\,320$ ($97.6\%$) inferred communities display a pronounced core structure.

\subsection{Core statistics}
\begin{figure}[!htbp]
\centering
\includegraphics[width=\linewidth]{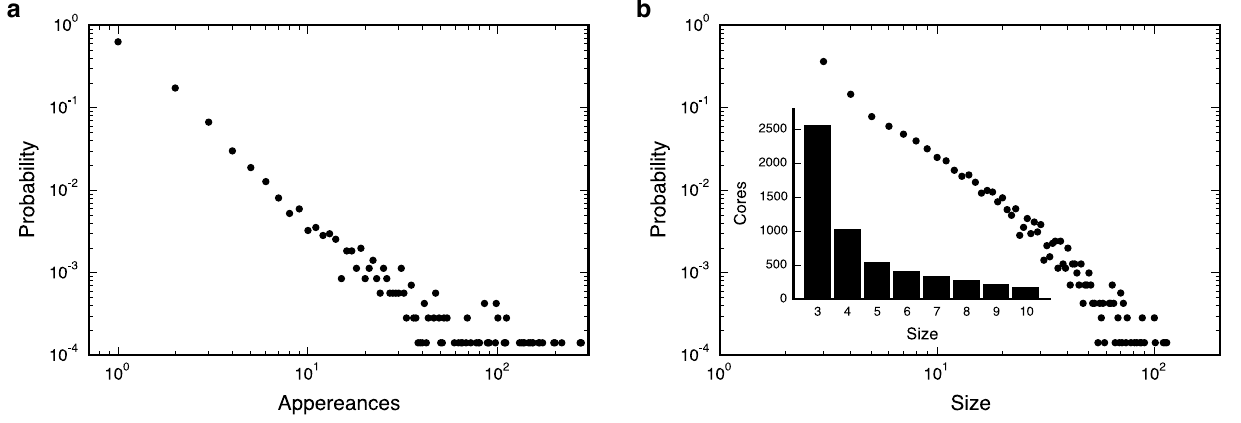}
\caption[Core statistics]{\small Core statistics, where we focus on cores of size greater than two. \textbf{a}, Probability distribution of the number of appearances per core. \textbf{b}, Size distribution. Inset shows raw numbers for specific sizes.} 
\label{fig:core_stat}
\end{figure}

Cores have a broad distribution of the number of appearances (Fig.~\ref{fig:core_stat}a), ranging from cores appearing only once to on average occurring multiple times per day.
The size distribution is also heavy-tailed (Fig.~\ref{fig:core_stat}b).

\begin{SCfigure}
\centering
\caption[Number of cores per individual]{\small Distribution of the number of cores per individual. Calculated for users that appear in frequently observed cores, i.e. cores that on average are observed more than once per month. A minority of individuals partake in more than 10 cores, with  the distribution being centered around a mean value of $\sim 5$. \vspace{20pt}}
\includegraphics[width=0.7\textwidth]{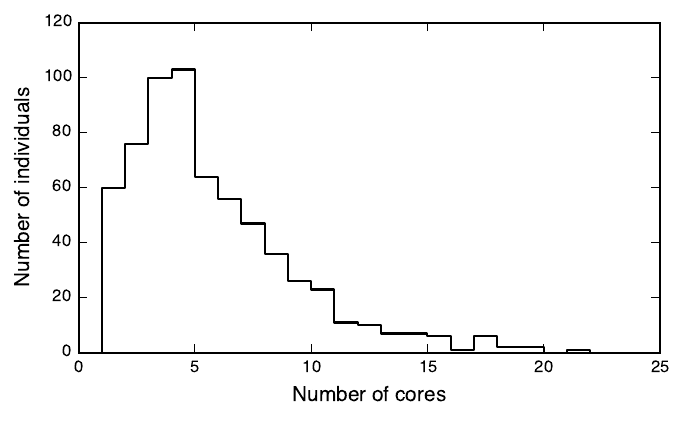}
\label{fig:cores_per_user}
\end{SCfigure}

To produce meaningful temporal statistics we henceforth focus on cores which on average are observed more than once per month,
this limits our focus to individuals that appear in these.
Fig.~\ref{fig:cores_per_user} shows the distribution of the number of cores each individual is part of; it is a broad distribution with a majority of individuals partaking in few cores while a small minority of users are extraordinary social and have more than 10 cores.  

\subsection{Subcores}
\begin{figure}[!htbp]
\centering	
\includegraphics[width=\linewidth]{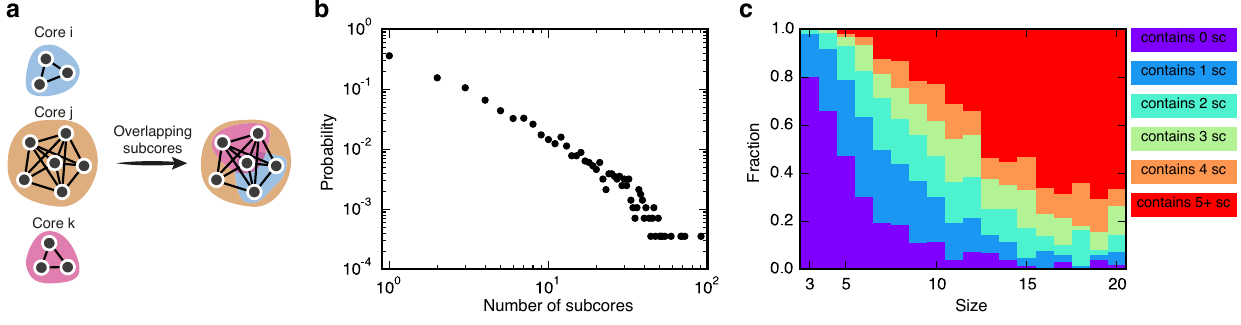}
\caption[Subcore structures and statistics]{\small Subcore structures and statistics. \textbf{a}, Illustration of overlapping and hierarchically stacked core structures. \textbf{b}, The distribution of the number of subcores contained within individual cores. \textbf{c}, Fraction of cores that contain $0$, $1$, $2$, $3$, $4$, and $5+$ subcores (sc) as function of core size.} 
\label{fig:subcore}
\end{figure}

Because cores span a wide range of sizes small cores can appear as subcores embedded within larger ones, see example in Fig.~\ref{fig:subcore}a.
In fact, our methodology allows for and identifies highly overlapping and hierarchically stacked structures.
We define a core to be a subcore if, and only if, it is fully contained in the larger one.
Figure~\ref{fig:subcore}b shows a broad distribution in the number of subcores that are contained within individual cores, with a majority of cores only containing one subcore, while other can contain more than ten.
There is of course a dependence on size, such that bigger cores have larger probability of containing more subcores.
Figure~\ref{fig:subcore}c quantifies this phenomenon by considering the fraction of subcores that cores of size $s$ contain.

\subsection{Work \& recreational cores}
\begin{figure}[!htbp]
\centering
\includegraphics[width=\linewidth]{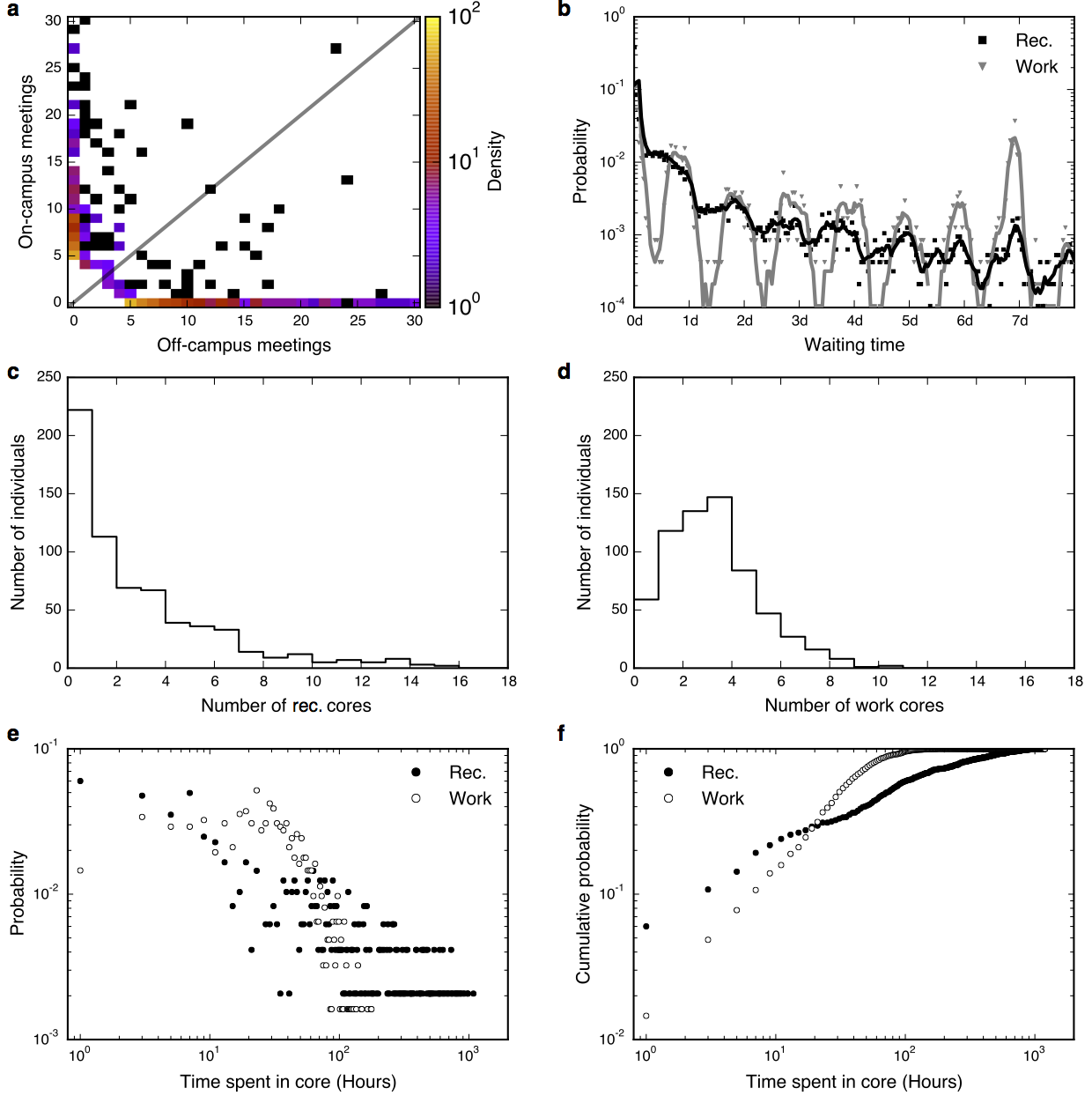}
\caption[Distinction between work \& recreational cores]{\small Distinction between work \& recreational cores. \textbf{a}, Excerpt of voting scheme depicting the number of on- and off-campus meetings for each core. Gray line splits the area into work (above) and recreation (below) categories. \textbf{b}, Inter event time distributions between consecutive meetings, aggregated across all cores of similar class (work/recreation). Events are hourly binned and full lines denote moving averages, calculated using 4-hour windows. \textbf{c}, Number of work cores per user. \textbf{d}, Number of recreational cores per user. \textbf{e}, The distributions of how much time individuals, in total, spend in cores with work and recreational context. The average individual spends approximately $38$ hours in a work setting, and $120$ hours in a recreational context during the period of study. \textbf{f}, Cumulative probability distribution of data shown in panel e.} 
\label{fig:work_play}
\end{figure}

In Sec. \ref{sec:on_off} we determined for each gathering whether it occurred on or off campus. 
For cores this distinction is not possible since cores consist of multiple gatherings that in principle can occur both on or off campus.
Instead we count the number of times each core appears on and off campus and perform a majority voting.
Cores that have a higher frequency of on-campus meetings are denoted as \textit{work} cores, while we call cores that appear more frequently off-campus as \textit{recreational} cores.
In case of a tie, we label the core as recreational.
Figure \ref{fig:work_play}a shows the voting schedule and indicates the split, while Fig.~\ref{fig:work_play}b shows the waiting time probability between consecutive meeting events.
For work cores the waiting time probability shows clear signs of daily and weekly patterns, suggesting that these cores may be driven by the class schedule.
Recreational cores on the other hand exhibit a more subtle pattern that slowly decays and is considerably higher during nighttimes---suggesting that two fundamentally different mechanisms drive the activity of the two groups.
Splitting the number of cores per individual (Fig.~\ref{fig:cores_per_user}) up into the two categories yields the results shown in Fig.~\ref{fig:work_play}c-d, where users have a broad degree of recreational cores, on average participating in $2.53$, while the number of work cores is more localized with an average value of $2.74$.
According to Fig.~\ref{fig:work_play}e-f individuals spend more time in recreational cores than work cores, clearly depicting that context has great influence on the properties of meetings.

\subsection{Meeting regularity}
\begin{figure}[!htbp]
\centering
\includegraphics[width=\linewidth]{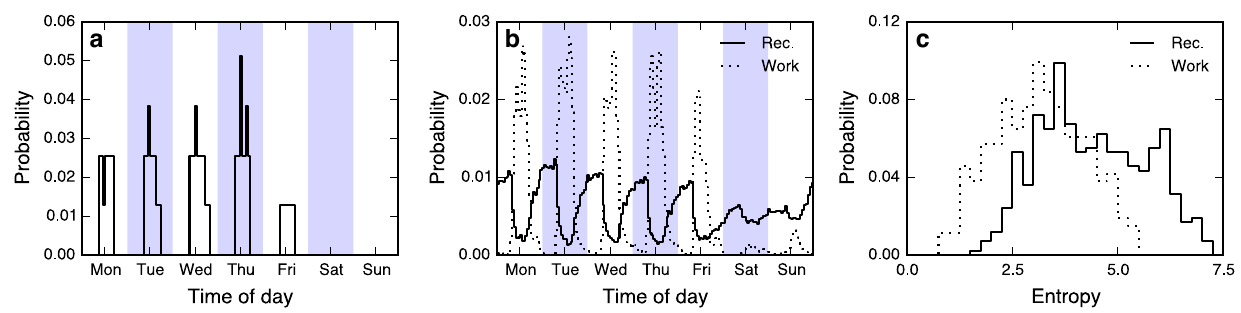}
\caption[Meeting regularity of cores]{\small Meeting regularity of cores. \textbf{a}, Example pattern for a single core, denoting the probability of observing the core. Data is aggregated across all weeks into 1-hour wide bins. \textbf{b}, Aggregated meeting patterns across all cores, showing the probability of observing work and recreational cores. \textbf{c}, Distributions of meeting time entropy calculated across all cores and divided up into work and recreation categories.}
\label{fig:temporal_patterns}
\end{figure}

Each core has a specific temporal pattern linked to it denoting the periods of time it has been present, see Fig.~\ref{fig:temporal_patterns}a for an example.
The information contained in each pattern can be quantified using Shannon entropy\cite{shannon1948entropy}, defined as
\begin{equation} \label{eq:shannon}
H=-\sum_{t} p_{t} \log_2 p_{t},
\end{equation}
where the sum runs over all temporal bins $t$ and $p_t$ is the probability of observing a specific core within given time-bin.
For each core we aggregate its meeting patterns across the full study duration into weekly 1-hour bins, then within each bin we calculate the probability of observing the core.
Entropy is calculated individually for cores using Eq. \ref{eq:shannon}.
According to Fig.~\ref{fig:temporal_patterns}b there is clear differences between the meeting patterns for work and recreational cores.
Further, the entropy distributions (Fig.~\ref{fig:temporal_patterns}c) reveal that recreational cores, on average, have higher entropy and thus lower meeting regularity---indicating that they do not meet within pre-scheduled temporal-bins.

\subsection{Ego viewpoint}
\begin{figure}[!htbp]
\centering
\includegraphics[width=\linewidth]{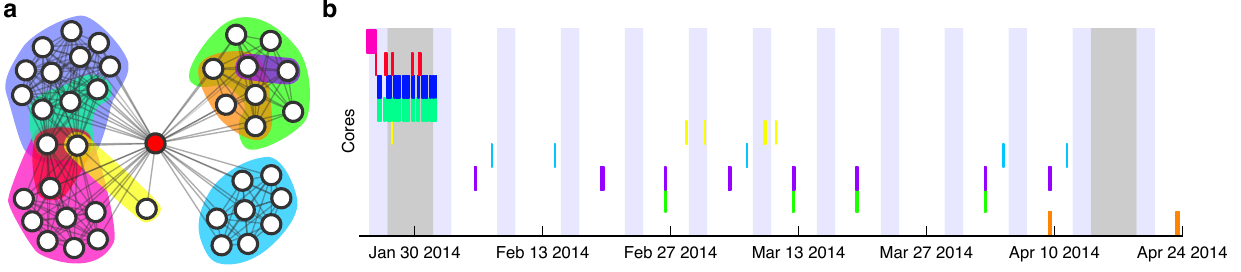}
\caption[Ego-centric perspective of cores]{\small Ego-centric perspective of cores for a representative individual. \textbf{a}, Network perspective, revealing overlapping and nested structures. \textbf{b}, Temporal dynamics of individual cores, colored accordingly.}
\label{fig:ego_perspective}
\end{figure}

So far we have mainly focused on the overall structural and dynamical features of cores, but we can reverse the perspective and look at cores from the perspective of individuals. 
From this perspective, each core provides a social context for the situation in which a person is embedded, whether it is in a work, recreation, or other relation.
Figure~\ref{fig:ego_perspective}a shows the involvement of a representative individual that is involved in multiple cores, producing hierarchically nested and overlapping structures.
The corresponding temporal pattern of cores (Figure \ref{fig:ego_perspective}b) reveals a complex behavior.

\section{Predicting behavior from routine}
In this section we describe the detailed analysis leading up to routine-based geographical location and social context prediction presented in the main text.

Song \emph{et al.}~have used entropy, an information theoretic measure in order to estimate the upper bound of the predictability of individuals' mobility patterns~\cite{song2010limits}.
We argue here, that in analogy to geospatial behavior, human social life can be described by a temporal sequence of `social states'. 
These can be used to quantify the predictability of social life.

Given a sequence of states for an individual $i$ we can define entropy in two ways.
First we can think of predictability in a temporally uncorrelated sense with entropy defined as
\begin{equation} \label{eq:S_unc}
S_i^{\text{unc}}=-\sum_j^{N_i} p_j \log_2 p_j,
\end{equation}
where $p_j$ is the probability of observing state $j$ and $N_i$ is the total number of states observed by person $i$.
Eq. \ref{eq:S_unc} captures the uncertainty of your location history without taking the order of visits into account, thus discarding information contained in the daily, weekly and monthly sequences of behavior.
A more sophisticated measure that includes temporal patterns is \emph{temporal entropy}:
\begin{equation}
S_i^{\text{temp}}=-\sum_{T_i' \subset T_i } p(T_i') \log_2 [p(T_i')],
\end{equation}
where $p(T_i')$ is the probability of finding a subsequence $T_i'$ in the trajectory $T_i$.
From the entropy one can estimate the upper bound of predictability ($\Pi_i$) by applying a limiting case of Fano's inequality\cite{fano1968transmission,song2010limits,jensen2010estimating}:
\begin{equation} \label{eq:pred}
S_i=H(\Pi_i)+(1-\Pi_i)\log_2(N-1),
\end{equation}
where $H(\Pi_i)=-\Pi_i \log_2(\Pi_i)-(1-\Pi_i)\log_2(1-\Pi_i)$.

\subsection{Comparing with previous studies} \label{sec:compare}
In the main manuscript (Fig. 3a) we show that our ability to predict the location of individuals has an upper bound of $71\%$ which is significantly lower than the $93\%$ reported by Song et al.~\cite{song2010limits}.
The main reasons behind this discrepancy are discussed below.

\subsubsection{Difference in populations}
Our study population is comprised of university students that (1) not necessary have a single home/nightly location, (2) have a rich free time which they can utilize to explore new locations, and (3) have multiple work locations due to classes being distributed across a large university campus.

In contrast, Song et al. studied a sample of $50\,000$ individuals selected from a total population of $\sim 10$ million anonymous mobile phone users.
The users were chosen according to (1) their travel patterns, where individuals had to visit more than two cell tower locations during the observational period of three months and (2) their average call frequency which had to be $\geq 0.5\text{ hour}^{-1}$, effectively selecting individuals with at minimum of 12 calls per day.

\subsubsection{Geospatial resolution}
\begin{figure}[!htbp]
\centering	
\includegraphics[width=\linewidth]{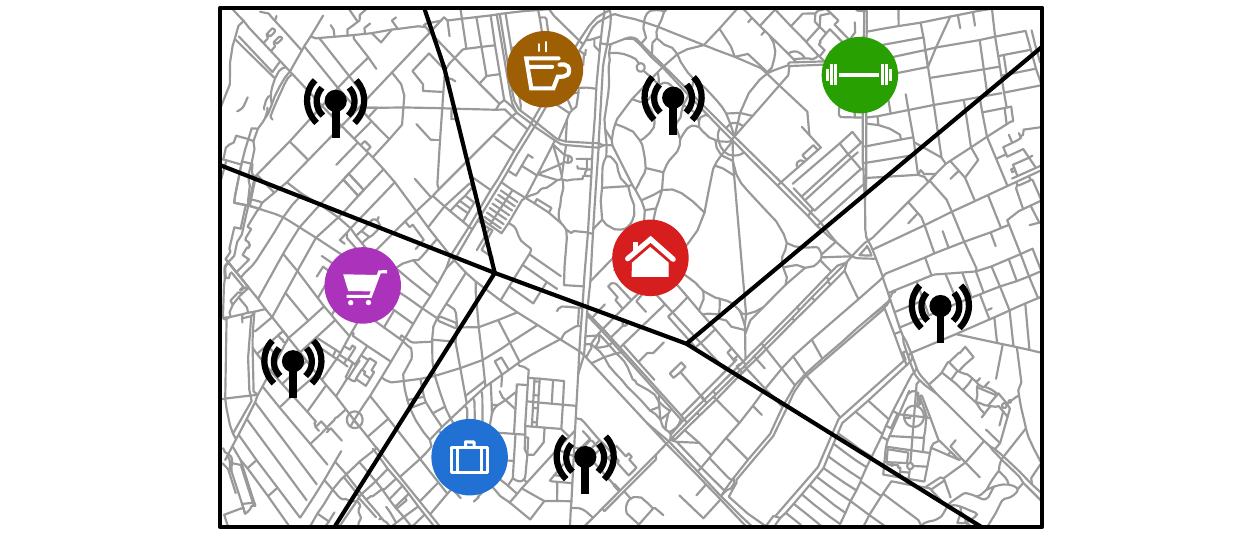}
\caption[Cell tower resolution]{\small Cell tower resolution of a city. Otherwise distinct places can be grouped together under the same cell tower, coarse graining the geospatial position and increasing the probability of guessing correct.} 
\label{fig:places}
\end{figure}

Previous studies have applied call detail records (CDR) as proxy for location, inferring the position of individuals depending on which cell tower their mobile phone is connected to during a call\cite{bagrow2012mesoscopic,gonzalez2008understanding,song2010limits}.
While the granularity of cell tower locations in cities is around $800$ meters it can be on the order of kilometers in more rural areas~\cite{trevisani2004cell}.
Figure \ref{fig:places} illustrates the effect of using cell tower data for positioning, as it can cluster otherwise distinct places together as one.
Our location data on the other hand has a typical accuracy below $60$ meters \cite{stopczynski2014measuring}, enabling a more accurate spatial estimation.

\subsubsection{Effects of binning}
Song et al. \cite{song2010limits} chose, because of data granularity, to segment their data into one hour time-windows.
Since our data has finer granularity, down to the minute scale, we can choose a different resolution, but which is optimal?

\begin{figure}[!htbp]
\centering	
\includegraphics[width=\linewidth]{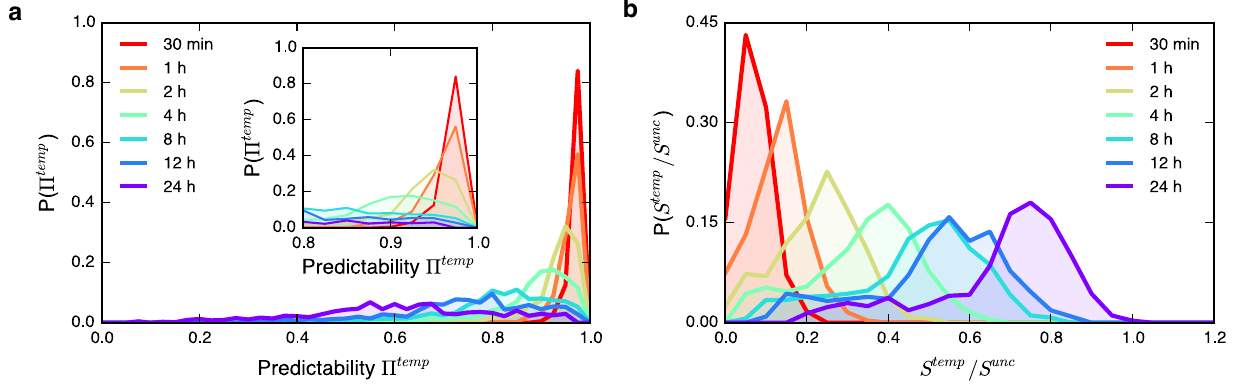}
\caption[Effects of binning on predictability]{\small Effects of binning on predictability bounds. \textbf{a}, Predictability as function of window size. Inset shows a close up for high values of predictability. Segmenting time into finer bins yields higher bounds of prediction. \textbf{b}, Ratio between temporal and uncorrelated entropy values. As the size of time-windows is narrowed the ratio shifts towards zero, indicating that $S^{temp}\ll S^{unc}$.} 
\label{fig:bin_size}
\end{figure}

We show in Fig. \ref{fig:bin_size}a that the finer we segment time, the better we are able to predict your location in the next time-bin.
By reducing temporal bin size it is possible to achieve arbitrarily high levels predictability because segmenting data into finer time-windows increases the number of bins, which in turn leads to respective states obtaining a higher frequency of visits.

The bin-size affects both temporal and uncorrelated entropy, however, so one hypothesis is that it is still meaningful to consider the ratio between the temporally uncorrelated entropy and the temporally correlated entropy. 
To investigate, we study on the ratio $S_i^{temp}/S_i^{unc}$, evaluated across all individuals $i$.
As is clear from Fig. \ref{fig:bin_size}b the ratio shifts towards lower values for small windows, revealing that the choice of bin width has a greater impact on temporal entropy.
This suggests that predictability is greatly influenced by the choice of time-window, as has also been noted by \cite{lin2012predictability,jensen2010estimating}.
In short, considering the ratio $S_i^{temp}/S_i^{unc}$ does not solve the binning problem. 
Ultimately this implies that the smaller bins we apply the better we are at predicting.
Because we currently are unaware of any timescale that is fundamentally descriptive of human behavior, we chose to work with temporal sequences in their natural form instead of segmenting them, predicting `next state' rather than `state of next time bin' (see Fig.~\ref{fig:vocabulary}).

\subsection{Data for prediction}
\subsubsection{Social vocabulary}
\begin{figure}[!htbp]
\centering
\includegraphics[width=\linewidth]{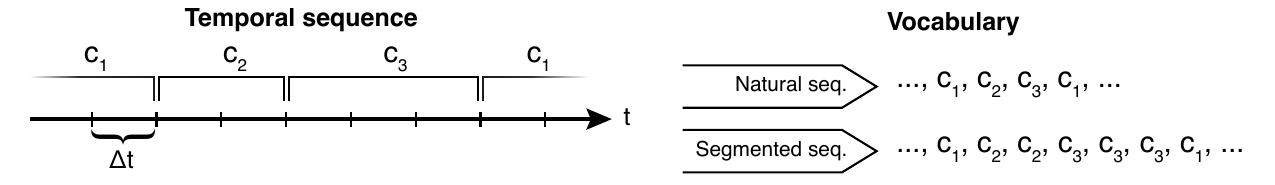}
\caption[Time-series of states]{\small Time-series of states. Left part of the figure presents social states as they naturally occur in a temporal sequence, with $\Delta t$ denoting a segmentation of the sequence within arbitrary sized time-windows. Right panel illustrates the difference in vocabularies, where the `natural' sequence focuses on the order of states, while `segmented' also weights states according to their duration. }
\label{fig:vocabulary}
\end{figure}

We describe the social life of an individual based on the context provided by cores, where we focus on individuals that appear in frequently observed cores (observed, on average, at least once per month).
In order to include the full social life of an individual we incorporate the context provided by dyads as well as the the information contained in infrequently observed cores.
Note that if a core/dyad is only observed once then we denote it as a `noise' state.
In addition we construct one supplementary context: `alone' denoting periods of time where an individual is not socially active.
Following Song et al.\cite{song2010limits} we construct a time-series of social contexts for each user.
However, as noted above, we do not segment the sequence into temporal bins, but keep it in its natural form, and predict `next state' (illustrated in Fig. \ref{fig:vocabulary}) for both geospatial and social prediction.

\subsubsection{Location vocabulary}
In addition to social context, we also collect geographical traces for each user, enabling us to reconstruct their mobility patterns.
To infer context from raw location traces we use the same definitions as Cuttone \emph{et al}.~\cite{cuttone2014inferring}, where a point of interest (POI) is a location of relevance for a person, such as home, work, or a cinema.
POIs are inferred by applying a density based clustering algorithm \cite{ester1996density}, with a density grouping distance of $60$ meters and requiring stops to consist of at least two samples, meaning that a person must have spent a minimum of $15$ minutes in the same location.


\subsubsection{Convergence of states}
\begin{figure}[!htbp]
\centering	
\includegraphics[width=\linewidth]{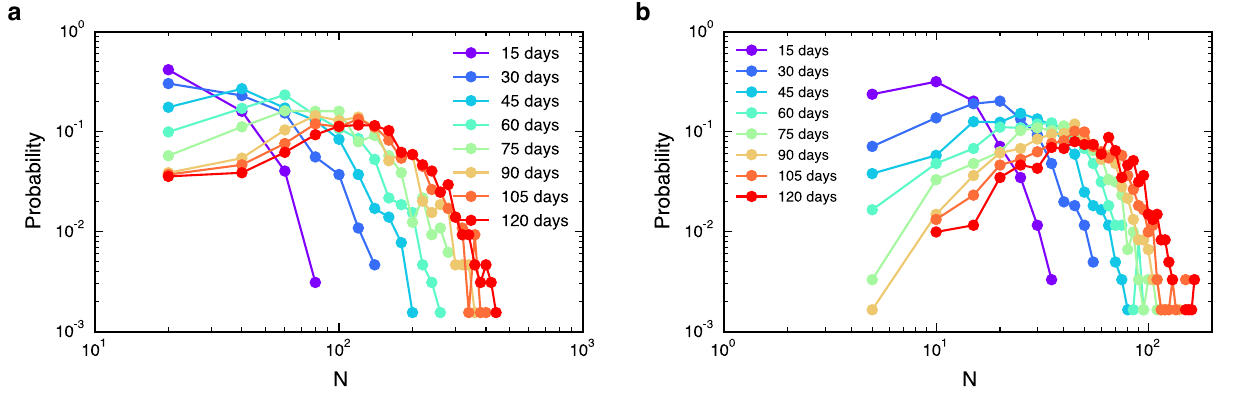}
\caption[Distribution of the number of distinct states]{\small Distribution of the number of distinct states within a time window. \textbf{a}, Convergence of the number of social social states, showing saturation after a time window of 90 days. \textbf{b}, Saturation of visited locations, also convergence after 90 days.} 
\label{fig:convergence}
\end{figure}

Figure \ref{fig:convergence} shows the distribution of the number of distinct social and geospatial states for increasing windows of time.
After 90 days both probability distributions converge, implying that the number of states visited by users is saturated, indicating that we can uncover a majority of states frequented by individuals. 
We, however, expect this saturation only to be meta-stable, because the social networks change across adulthood~\cite{wrzus2013social}.


\subsection{Prediction}
\begin{figure}[!htbp]
\centering	
\includegraphics[width=\linewidth]{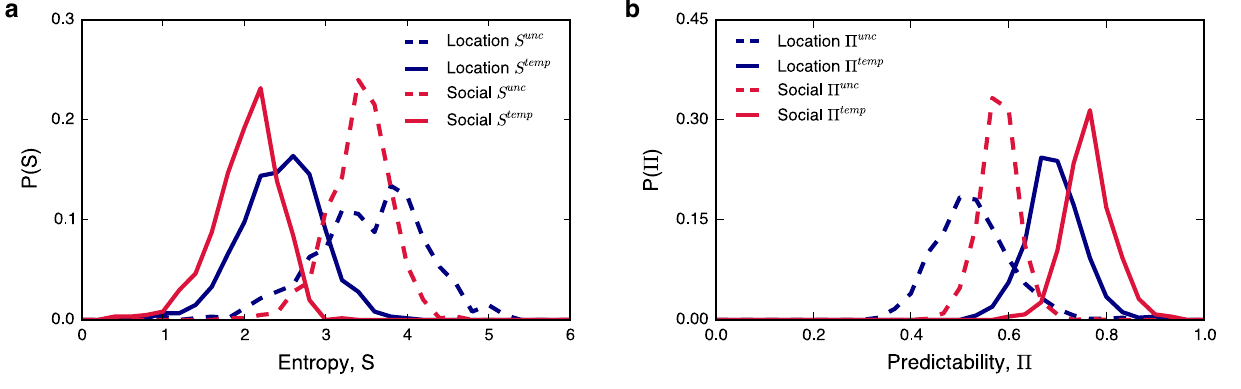}
\caption[Distribution of entropy and predictability]{\small Probability distributions of entropy and predictability. \textbf{a}, Distributions of temporal and uncorrelated entropy for location and social behavior. As expected, temporal patterns contain more information than just frequency of visits, and hence have a lower entropy. \textbf{b}, Predictability distributions for uncorrelated and temporal patterns. On average, our social behavior is more predictable than our geospatial behavior.} 
\label{fig:predictability}
\end{figure}

Following Eq.~\ref{eq:S_unc}-\ref{eq:pred} we first calculate the respective entropies of the behavioral patterns for each individual.
We show in Fig.~\ref{fig:predictability}a that our social patterns have lower entropy than our mobility.
The figure shows that an average person approximately occupies $2^2\approx 4 $ social states and $2.5^2\approx 6.25$ location distinct states.
We also find that humans are potentially more predictable based on their social contexts than their past locations, see Fig. \ref{fig:predictability}b. 
Previous studies have found higher levels of predictability\cite{song2010limits,lu2012predictability}, see Sec. \ref{sec:compare} for a full discussion.

\begin{figure}[!htbp]
\centering	
\includegraphics[width=\linewidth]{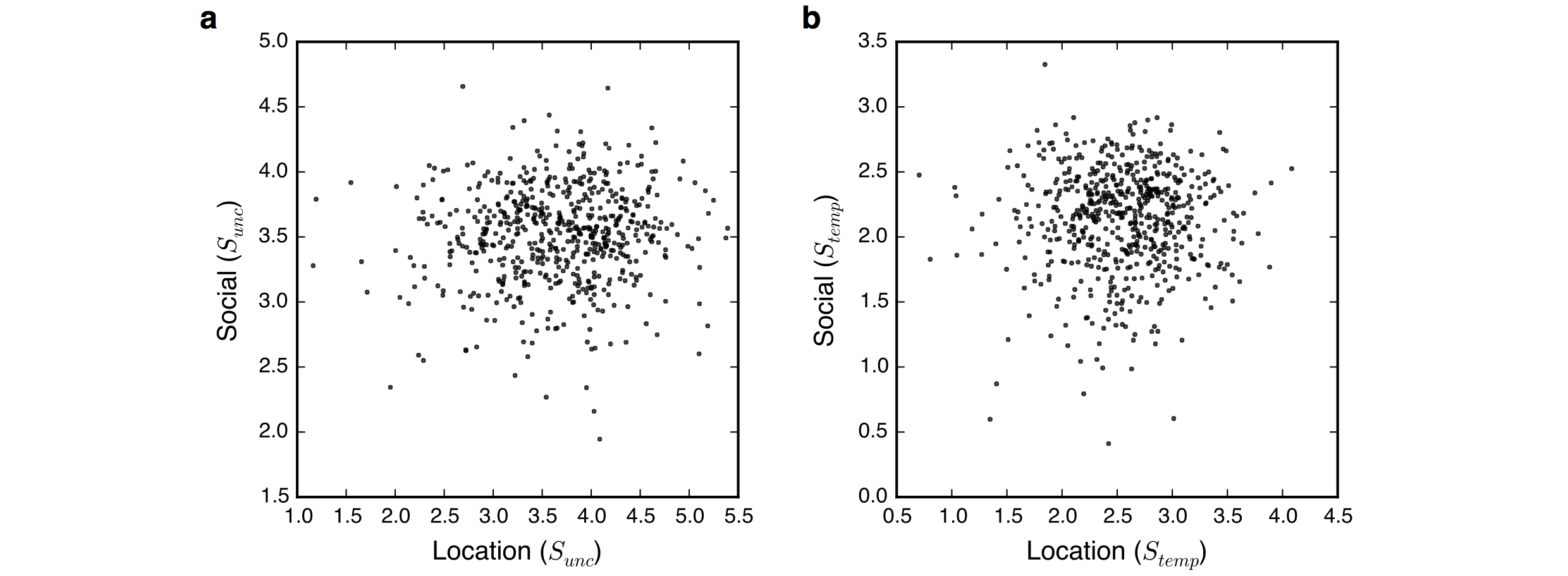}
\caption[Correlation between social and location entropy]{\small Correlation between social and location entropy. \textbf{a}, Mutual dependence between uncorrelated entropy for social and location states per individual. \textbf{b}, Correlation between temporal entropies for social and location states.} 
\label{fig:correlation}
\end{figure}

Fig. \ref{fig:correlation}a shows the interrelation between $S^{unc}_{social}$ and $S^{unc}_{location}$, surprisingly there is no correlation between the two measures (Spearman correlation $\rho = 0.053 $, $p$-value $= 0.191$), indicating that humans can be highly predictable in a social sense but very unpredictable location-wise and vice versa. 
A similar lack of correlation is observed for $S^{temp}$ (Spearman correlation $\rho = -0.008 $, $p$-value $= 0.84$), see Fig. \ref{fig:correlation}b.

\subsection{Temporal aspect of predictability}
\begin{figure}[!htbp]
\centering	
\includegraphics[width=\linewidth]{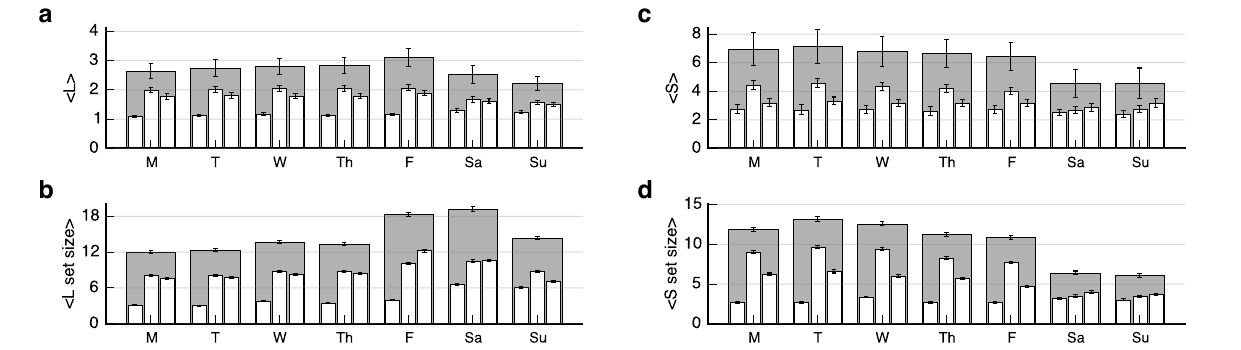}
\caption[Temporal aspects of predictability]{\small Nested histograms showing the temporal aspects of predictability. Binned using daily and 8-hour intervals (12 am - 8 am, 8 am - 4 pm, 4 pm - 12 am), outer bars (gray) denote days while inner bars (white) denote 8-hour windows. Bars do not necessary add up, because one can have an overlap of states between the 8-hour bins. All values are averaged across the student population. \textbf{a}, Number of average observed locations per bin. \textbf{b}, Total number of visited distinct locations. \textbf{c}, Average number of social states per time-bin. \textbf{d}, Total number of distinct social states.} 
\label{fig:states}
\end{figure}

According to Fig. \ref{fig:correlation} there is no correlation between social and geospatial aspects of human life, but this is measured in terms of overall dependence.
In real life we have varying degrees of predictability, a simple way of visualizing this is to look at the number of states as function of time. 
According to Fig. \ref{fig:states}a we have a low number of location states during nights (resulting in low entropy) because we mainly sleep at a single location, while during days and evenings our entropy is higher, in part because because we occupy more states.
Note that there is very little overlap between locations visited in each 8-hour bin, e.g. morning locations are different from day locations.
On Fridays we visit more locations than any other day, while during weekends we are more stationary.
If we, however, consider the total number of distinct visited places (Fig. \ref{fig:states}b) we see that Fridays and Saturdays are special because those days are used to explore new locations.
Therefore, predicting location during weekends based on routine is more difficult, since we have higher entropy during these periods.
Our social behavior (Fig. \ref{fig:states}c) resembles our mobility, where we socialize mainly during the day and less during the night.
Weekends are again special, interestingly we here observe a drop in in the number of social states, because we are not required to go to work or school. 
Fig. \ref{fig:states}d shows that the number of social states decreases during weekends, meaning our participants reserve weekends to socialize with a few selected friends.

Here, it is important to realize that we only observe social interactions among participants of the experiment. 
Wile the geospatial data is sampled evenly over the observation period, social interactions might have a potential bias, as it is possible to go out with non-university friends on Fridays and weekends.
We address this issue, in Fig. 5b (main manuscript), by quantifying the geo-spatial behavior of cores.
Whenever a core is present we pool together the geographic locations of its individuals, bin the resulting list of (latitude, longitude)-points in 0.01 x 0.01 cells, and quantify the behavior using uncorrelated entropy.
Using 0.1 x 0.1 and 0.001 x 0.001 cells produces comparative results. 
Displayed in the figure is the average entropy, averaged over all cores and binned in 8-hours bins (as described above).

\section{Social prediction}
It has previously been shown that spatial behavior between individuals that share a social tie is correlated \cite{crandall2010inferring,cho2011friendship,wang2011human,de2013interdependence}.
But the onset of co-presence lacks a temporal signature, so while the spatial traces overlap it is not generally possible to specify exactly when a friend is predictive for an individual's behavior.
Cores, however, do provide such context. 
An incomplete set of members provides a clue that a social interaction is about to occur (i.e.~the final group member is about to arrive).

We test this concept on cores of size three; thus provided we observe two members we measure the probability of the last member arriving within the next hour.
To avoid testing the hypothesis on scheduled meetings we focus on weekday nights (6 pm - 8 am) and weekends.
This is the period where routine driven prediction is at its weakest.
Further, in order to avoid circularity, we evaluate the hypothesis on a test month (May 2014), during which we have not identified gatherings. 

\subsection{Null models}
For evaluation purposes we compare cores to two reference models, both generated from real world data.
We segment interactions into undirected and unweighted daily graphs, see Fig. \ref{fig:social_prediction}a.
The first null model, we which we denote \textit{random}, constructs reference groups by randomly drawing nodes from each daily graph.
The second model utilizes a breadth-first search in the daily graph to create reference groups starting from a randomly chosen seed node by searching its local neighborhood.
A new seed node is chosen if the search is restricted to components with fewer than three members.
\textit{BFS} is a strict null model and requires all nodes to have shared a physical interaction.
We disregard reference groups if they happen to be identical to a core.

\subsection{Comparison}
We test the hypothesis on a sample of $340$ frequently occurring cores and $10\,000$ reference groups for $n=100$ independent trials.
For $33$ cores we never observe an incomplete set of members during the month of May, therefore they cannot be used for prediction and are disregarded.

\begin{figure}[!htbp]
\centering	
\includegraphics[width=\linewidth]{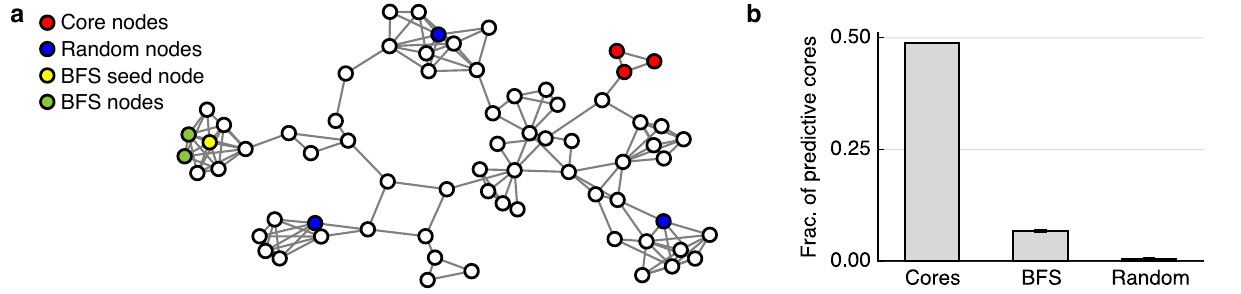}
\caption[Social prediction]{\small Social prediction. \textbf{a}, Daily graph of interactions, illustrating cores and construction of null models. \textbf{b} Percentage of socially predictive cores within each category. A group is predictive if it at least once has correctly predicted the arrival of the missing individual. For the reference models errorbars are calculated across $n=100$ independent trials.} 
\label{fig:social_prediction}
\end{figure}

As shown in Fig. \ref{fig:social_prediction}b approximately $50\%$ of cores are predictive, in stark comparison to BFS reference groups, with the random null model performing even worse.

\section{A dynamic random geometric graph model}
In order to investigate the statistical significance of the results found in the empirical data, we create a null model to simulate full network dynamics. 

The simplest candidate for a model system is based on random geometric graphs (RGGs) \cite{penrose2003random} which incorporate random walks. 
We model a scenario where individuals move randomly and connect to each other when they are physically proximate. 
This model has three parameters. 
The random walker step size $\ell$, population density $\rho$ and contact radius $r$ (of which only two are free after rescaling).
In the current setting, it is natural to choose a scale for this model that lets us interpret the results in units that correspond to our actual real-world system, thus we connect two links if their distance is less than $r=10$ meters, corresponding to the typical Bluetooth range used in the main text.

\subsection{Single time-slice RGG parameters}
Random geometric graphs have been used as models for complex networks in the literature, see Ref.~\cite{barthelemy2011spatial} for an overview. 
In a RGG, we expect to find a single giant connected component when $r$ is much greater than the typical distance between neighboring nodes. When $r$ is much smaller than the typical distance between neighboring nodes, the random geometric graph has many isolated components with the component size distribution approximately exponential \cite{PhysRevE.66.016121}. 
Near the critical threshold $r_c$, the distribution of component sizes follows a power-law \cite{PhysRevE.66.016121}. 

\begin{figure}
	\centering
	\includegraphics[width=\hsize]{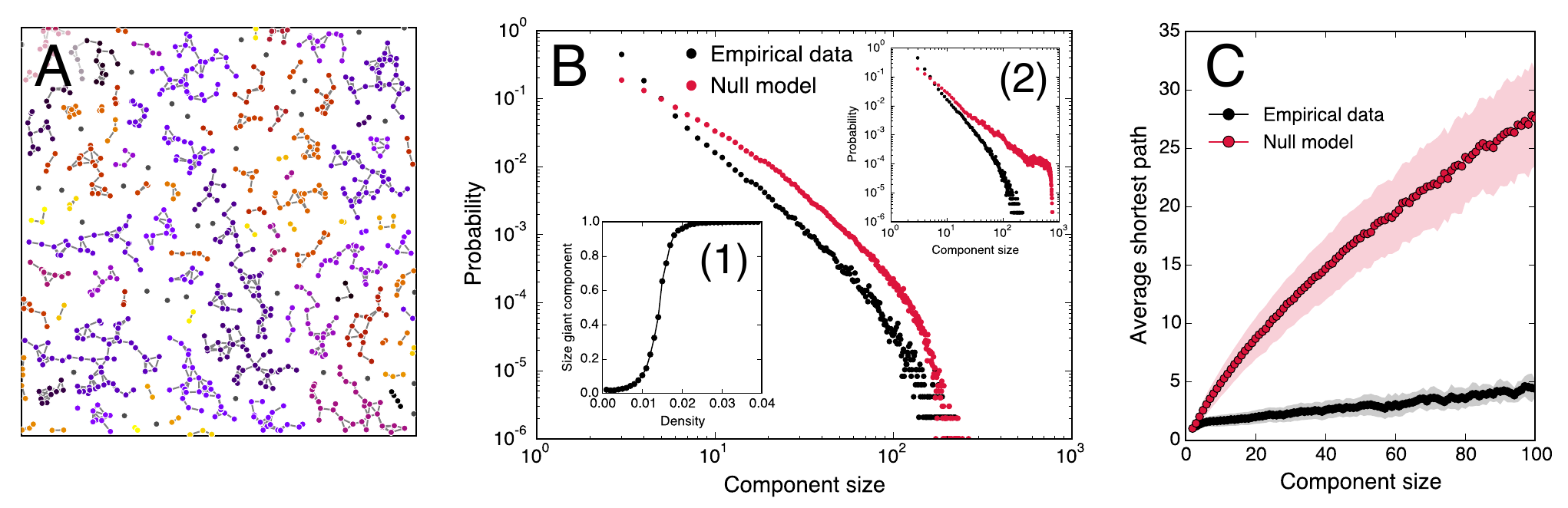}
	\caption[The random geometric graph model]{
The RGG model (parameters as described in the main text). \textbf{A}. A snapshot of the model near the percolation threshold; notice how components do not form dense clique-like structures. 
\textbf{B}. The normalized distribution of component sizes in many realizations of the RGG model (red) and as observed in the empirical data (black). 
When the random geometric graph is near the percolation threshold, both systems display power distributions of component sizes. 
Inset~(1) shows the average largest Connected Component size (normalized to total number of nodes) as a function of density. 
Inset~(2) shows the real and model component size distribution at the critical point. 
\textbf{C}. The Average shortest path length inside components in the RGG (red). Shaded areas denote standard deviation. The Average shortest path length inside components in real data (black). Notice how the average shortest path length tends to be significantly larger in the sparse, spatially extended components generated by the RGG model. \label{fig:static}}
\end{figure}

For our purposes, RGGs are interesting as a model for our real-world system around $r_c$ because gatherings in the real-world network display a power-law distribution of sizes, as suggested by the distribution of gathering sizes displayed in Figure~\ref{fig:gath_stat}. 
Staying with parameters chosen to correspond to the empirical data, we construct a random geometric graph with $n=800$ nodes distributed randomly in a $L\times L$ square, corresponding to density $\rho = n/L^2$, see Fig.~\ref{fig:static}A for an illustration, to remain consistent with the literature, we use open boundary conditions when forming components \cite{PhysRevE.66.016121}.

Using the interpretable `physical' parameter choices described above, we can estimate $\rho$ based on the critical threshold for RGGs reported in the literature \cite{PhysRevE.66.016121, 0305-4470-33-42-104}. 
This yields $\rho = 0.014 \,\,\textrm{nodes}/m^2$, in accordance with our numerical simulations, see Fig.~\ref{fig:static}B (inset~1). 
The resulting distribution is shown in Fig.~\ref{fig:static}B (inset~2). At the critical point, the distribution of component sizes in the RGG model is broader than the distribution found in real-world data and with a noticeable cut-off due to the finite size effects. 
By reducing the model's density by a small amount (to $\rho \approx 0.01$), we find that the distribution of component sizes in the RGG model can be tuned to match the real-world component size distribution reasonably well. 
(Details on distribution matching can be found below.) 
The final single time-slice RGG component size distribution for the `physical' parameters defined above is shown alongside its real-world counterpart in the main panel of Fig.~\ref{fig:static}B

\subsection{Model versus the real world}
While the distributions of component sizes in the model is qualitatively similar to what we find in the real data, it is important for our understanding of the model's limitations to point out that there are significant differences between the RGG and real-world data. 

Firstly, the spatial distribution of node-locations is \emph{very} different, from $\rho = 0.01 \,\,\textrm{nodes}/m^2$ and $n = 800$ nodes and resulting $L \approx 283 \,m$. 
In contrast, the real data-set finds individuals distributed in small clusters throughout the entire greater Copenhagen region ($L_\textrm{real}\sim20\,000$ meters, cf.~Ref~\cite{stopczynski2014measuring}).

Secondly, while both the real data and the RGGs have power law distributed component sizes, see Figure~\ref{fig:static}B, the internal structure of components in the two models are quite different. 
In Figure \ref{fig:static}A we show a realization of a RGG for the final parameter values discussed above.
Nodes that are within a radius of $r=10\,m$ are connected by a link. 
While more clustered than components in a Erd\H{o}s-R\'{e}nyi graph \cite{PhysRevE.66.016121}, the component we find in the RGG model are not clique-like as we observe in the empirical data (cf.~main text Figure 1). 
Instead of being clique-like, the RGG components have long average internal paths and significant spatial extension (the distance between extremal points often representing a significant fraction of the side-length $L$).
One way of quantifying the differences between the empirical components and the RGG components is considering the average shortest path length inside components in the two models. 
This is what we explore in Figure~\ref{fig:static}C, where we quantify how paths are significantly longer in the random geometric graph than what we observe in the more clique-like components observed in the empirical data. 

\subsection{A dynamic RGG model}
So far we have only considered static random geometric graphs---describing how to generate a single time-slice with statistics that match our real-world data.
We now implement graph dynamics by letting each node move according to a random walk; we use periodic boundary conditions for the random walkers.
We implement the random walk as follows. 
At time step $t$, node $i$ with location $(x_i(t),y_i(t))$ updates its position by selecting an angle $\theta_i(t) \in [0,2\pi)$ at random and moving with step length $\ell$ in that direction, ending up in $(x_i(t+1),y_i(t+1)) = (x_i(t) + \ell \cos[\theta_i(t)] ,y_i(t) + \ell \sin[\theta_i(t)])$, modulo the periodic boundary conditions.

\begin{figure}
	\centering
		\includegraphics[width=\hsize]{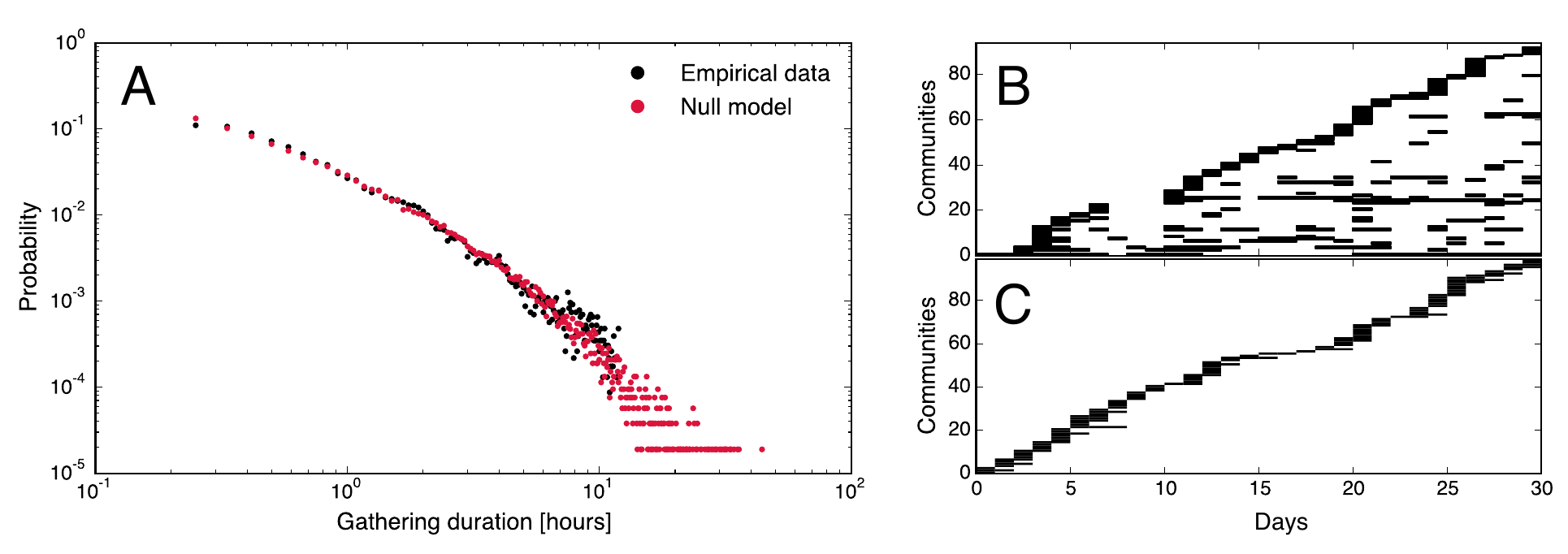}
	\caption[Dynamic random geometric graph model]{\textbf{A}. Distribution of gathering lifetimes for parameter values defined in the text.  \textbf{B} A example of participation in cores in the real world data (from main text Fig.~3d). \textbf{C} An example of participation in cores in the dynamic RGG model. Notice how the dynamic RGG does not have recurring gatherings. \label{fig:steps}}
\end{figure}

We choose step size $\ell$ to produce gathering lifetimes corresponding to the empirical distribution, given the choices of values for $n,r,\rho$ listed above. 
We match up the two distributions by iterating over values of $\ell$ and pick the step size which minimizes the Kullback-Leibler (KL) divergence \cite{kullback1951information} between the two probability distributions.
We find that $\ell=0.65$ (i.e. 0.65$m$ in the physical interpretation) produces lifetimes with a distribution of gathering lifetimes similar to that of the empirical system, see Figure~\ref{fig:steps}A for the two distributions.

Thus, quite remarkably, the very simple dynamic RGG model based on random walks for each node is able to reproduce, not only the the distribution of gathering sizes (Figure~\ref{fig:static}B), but also the distribution of gathering lifetimes.
This implies that in terms of \emph{gatherings} this very simple model is able to produce behavior quite similar to what we observe in the empirical data---at least in terms of distributions.

\subsection{Random walks in 2D}
Let us quickly recapitulate the central results on random walks in 2D\footnote{Random walks in 2D are well understood, see e.g.~\href{http://mathworld.wolfram.com/RandomWalk2-Dimensional.html/}{http://mathworld.wolfram.com/RandomWalk2-Dimensional.html/} for a nice overview}. 
The first key result is that the the average displacement of a random walker is zero.
This result simply means that if we choose a point $(x_0,y_0)$ and start an ensemble of random walkers there, each one moving according to the rules outlined above, their \emph{average} location will remain $(x_0,y_0)$ independent of the number of steps. 
The second result is that after $T$ steps, the root mean square displacement $d_\textrm{rms}$ of a walker with stepsize $\ell$ is $d_\textrm{rms} = \ell \sqrt{T}$. 

Taken together these two results characterize the behavior of individual random walkers. 
A walker does not remain at its starting point, but tends to drift, on average traveling some distance $d$ in $T \approx (d/\ell)^2$ steps. 
The fact that the average displacement is zero simply implies that a walker is equally likely to drift in any direction.
Thus if we find multiple walkers close to each other---as is the case for a gathering in our model---we expect the gathering to disappear after some time $T$, as the members each drift in a random direction.

\subsection{The dynamic RGG model does not generate recurring meetings}
So far, we have shown that the dynamic RGG model is able to reproduce the distributional properties of our real-world data. 
That is, for a single time-slice, the model can generate component size distributions that are quite similar to what we observe in the empirical data.
When we add a dynamic component to the model, allowing each node to move according to a random work, we see that the real-world distribution of gathering life-times can also be matched by this simple model.

The dynamic RGG model, however, does not generate cores\footnote{Technically, a given group of people only meeting a single time is both a gathering and a core (a core which occurs only once). Here, we use core in the sense that we do in the main paper---to signify recurring gatherings consist of 3 or more individuals and meet more than once per month on average}.
Recall that cores are \emph{recurring} gatherings, where a specific group of individuals meets repeatedly over time. 
Given the known dynamics of random walkers summarized above, the long-term picture finds each random walker drifting according to its own trajectory. 
Since there are no correlations between individual trajectories, the probability that the members of any gathering with more than two nodes will be in close proximity again, on the time-scale we are studying here (e.g.~simulating $\sim$ one month of data with parameters as above), is vanishing.
Thus, we do not expect to see cores occurring more than once in the dynamic RGG model. 

This is precisely what we observe when we analyze the model data. 
Figure\,\,\ref{fig:steps}B is a similar to the main text's Figure 3d, but this time showing cores with any number of occurrences (not just the recurring gatherings), displaying how a representative individual participates in her/his cores as a function of time. 
Time runs along the $x$-axis and each vertical line represents an activation of that core.
In the main text, we call this the path through the cores an individual's `social trajectory'.
In Figure\,\,\ref{fig:steps}B, we have generated a social trajectory for a representative node in the dynamic RGG model.
The differences in core participation patterns illustrates the argument above nicely.
Over the same time-period, the dynamic RGG model-node in question participates in nearly a similar number of social contexts, but all of them occurring only once; sometimes gatherings in the model can last multiple days.
In the real-world data, however, a pattern of recurring meetings is evident. It is based on these patterns that we use the cores to make predictions.

Thus, it is clear that due to the lack of recurring cores, the dynamic RGG model does not have predictability in the senses that we discuss in the main text.
Let us be more explicit about this point.
Recall that in the main text, we define predictability in two senses. 
Below, we discuss each one in detail.
\begin{itemize}
\item Firstly, we discuss what we call the `social unit property'. 
The basic idea behind the social unit property is that, cores represent meetings between $n>2$ individuals, characterized by the fact that all $n$ members are usually present when the core is active. 
Therefore an observation of an incomplete set of core members implies that the remaining members will arrive shortly, a fact which can be used for prediction.
It is trivial to see that cores in the dynamic RGG model do not possess the social unit property, simply because cores typically only occur once. 
\item Secondly, we measure predictability as defined in Ref.~\cite{song2010limits} (see main text for details).
The calculation for predictability in this sense depends on compressing the social trajectory. 
As is clear from Figure~\ref{fig:steps}B, the social trajectory for a node in the dynamic RGG model simply grows over time, continually introducing new gatherings that only occur once. 
This implies that the entropy for a node continues to grow as a function of time.
The fact that the entropy keeps growing should not surprise us, given that social structures arising in the simple RGG model are based on a random process. 
The fact that the entropy is unbounded implies that there is zero predictability in the Ref.~\cite{song2010limits} sense as well.
\end{itemize}

As noted earlier, RGG's have been used as models for social systems in the literature.
Specifically, Refs.~\cite{PhysRevLett.96.088702,starnini2013modeling} have explored modeling social networks using variations of the simple RGG model explored above, also finding that distributional properties of empirical data can be matched quite well.
In both cases the walkers `slow down' (decrease step size) upon forming links with other agents.
This social slow-down results in denser (more realistic) gatherings, but due to the fundamentally random nature of walkers in each model, neither of these model display cores that appear multiple times.
Thus, just like the simple dynamic RGG, these models do a surprisingly good job of simulating behavioral data for a single day (creates gatherings), but none of them capture behavior across weeks and months (unable to create realistic cores).  

\subsection{Summary}
In summary, the random geometric graph with random walkers provides an interesting model which is able to reproduces distributional properties of component sizes as well as component life-times quite accurately.
Therefore this model is an excellent point of comparison, when we want to understand the significance of our findings regarding the empirical dynamic networks. 

There are some important differences between the model and the empirical data.
When we look at the RGG components, for example, we find that their spatial configuration is quite different from what we observe in real data.
In order to reproduce the distributional properties, all model components must be packed in a very small area (a square with side length $L\approx 280$ meters) and the model component are sparser with much longer path-lengths than components in the real-world data.

The most important difference between the empirical data and the simple RGG model, however, is that the latter does not display any meaningful core structure.
As we have shown above, this implies that the predictability we observe in the empirical dataset is not replicated by the dynamic RGG model and can be considered a real effect expressed in the data.
\section{Coordination of meetings} \label{sec:coordination}
\begin{figure}[!h]
\centering	
\includegraphics[width=\linewidth]{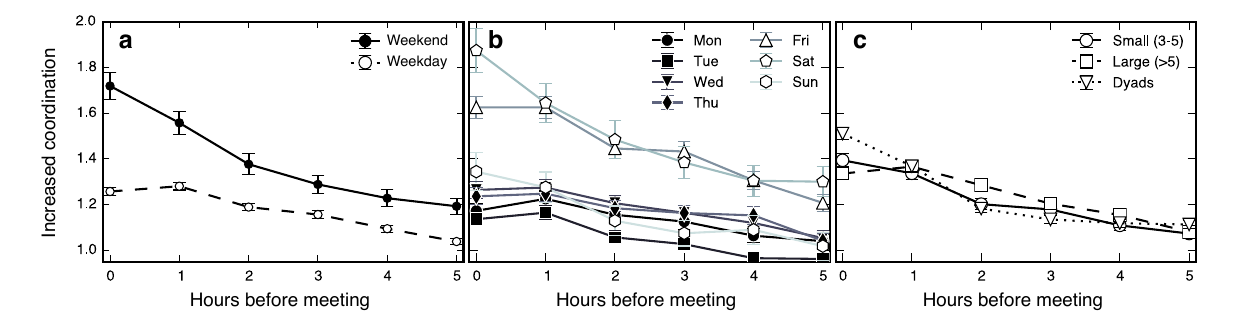}
\caption[Increased coordination prior to a meeting]{\small Increased coordination prior to a meeting. Calculated for all nodes participating in the first hour of a gathering. \textbf{a}, Required amounts of coordination between nodes, depending on when the gatherings meets. More coordination is required to organize meetings during weekends (Friday 4 pm - Sunday) than during weekdays (Monday - Friday 4 pm). \textbf{b}, Further sub-dividing the categories from panel a reveals that Fridays and Saturdays are special. \textbf{c}, Effect of size of the meeting on coordination. On average it requires equal amounts of coordination, per person, to organize a meeting, independent of the size of the group.} 
\label{fig:coordination}
\end{figure}

This section describes how we calculate the amount of coordination leading up to gatherings.
Prior to a meeting individuals might need to coordinate about when and where to meet.
This coordination can be conveyed through various means: (1) individuals can organize in real time through electronic means, such as online social networks and mobile phones, (2) they can verbally have scheduled meetings beforehand, i.e. at previous gatherings, (3) or attend routine driven pre-scheduled meetings, arranged by an institution, e.g. the university.
We consider the coordination in the hours leading up to a meeting, measured in terms of increased calling and texting activity.
Because calling frequencies change over the course of a day, and because individuals can have fundamentally distinct calling patterns \cite{onnela2007structure,jo2012circadian,stopczynski2014measuring}, we compare activity leading up to a meeting to hour-by-hour dynamic individual baselines, defined as
\begin{equation}
c_t = \frac{1}{N} \sum_{n=1}^N \frac{a_t^n}{\widetilde{a_t^n}},
\end{equation}
where $N$ it the number of individuals participating in the gathering, $a_t^n$ is the activity of person $n$, and $\widetilde{a_t^n}$ is the baseline activity. 
The equation denotes increased coordination levels $t$ hours before a meeting.
Because gatherings have a broad distribution of lifetimes (Fig. \ref{fig:gath_stat}), we restrict the calculation to individuals that participate in the first hour of the meeting. 
According to Fig. \ref{fig:coordination}a-b, meetings during the weekend require more coordination than meeting during weekdays. 
This implies that weekend behavior is less scheduled, emphasizing the problem of predicting behavior using traditional routine-based measures.
In addition, Fig. \ref{fig:coordination}c reveals that meetings, independently of size, require the same amount of coordination per person, illustrating the validity of Fig. \ref{fig:coordination}a for varying gathering sizes.

\newpage
\bibliography{social}

\begin{thebibliography}{10}
\expandafter\ifx\csname url\endcsname\relax
  \def\url#1{\texttt{#1}}\fi
\expandafter\ifx\csname urlprefix\endcsname\relax\def\urlprefix{URL }\fi
\providecommand{\bibinfo}[2]{#2}
\providecommand{\eprint}[2][]{\url{#2}}

\bibitem{simmel1950quantitative}
\bibinfo{author}{Simmel, G.}
\newblock \bibinfo{title}{Quantitative aspects of the group}.
\newblock \emph{\bibinfo{journal}{The Sociology of Georg Simmel}}
  \bibinfo{pages}{87--177} (\bibinfo{year}{1950}).

\bibitem{palla2007quantifying}
\bibinfo{author}{Palla, G.}, \bibinfo{author}{Barab{\'a}si, A.-L.} \&
  \bibinfo{author}{Vicsek, T.}
\newblock \bibinfo{title}{Quantifying social group evolution}.
\newblock \emph{\bibinfo{journal}{Nature}} \textbf{\bibinfo{volume}{446}},
  \bibinfo{pages}{664--667} (\bibinfo{year}{2007}).

\bibitem{ahn2010link}
\bibinfo{author}{Ahn, Y.-Y.}, \bibinfo{author}{Bagrow, J.~P.} \&
  \bibinfo{author}{Lehmann, S.}
\newblock \bibinfo{title}{Link communities reveal multiscale complexity in
  networks}.
\newblock \emph{\bibinfo{journal}{Nature}} \textbf{\bibinfo{volume}{466}},
  \bibinfo{pages}{761--764} (\bibinfo{year}{2010}).

\bibitem{holme2012temporal}
\bibinfo{author}{Holme, P.} \& \bibinfo{author}{Saram{\"a}ki, J.}
\newblock \bibinfo{title}{Temporal networks}.
\newblock \emph{\bibinfo{journal}{Physics reports}}
  \textbf{\bibinfo{volume}{519}}, \bibinfo{pages}{97--125}
  (\bibinfo{year}{2012}).

\bibitem{starnini2013modeling}
\bibinfo{author}{Starnini, M.}, \bibinfo{author}{Baronchelli, A.} \&
  \bibinfo{author}{Pastor-Satorras, R.}
\newblock \bibinfo{title}{Modeling human dynamics of face-to-face interaction
  networks}.
\newblock \emph{\bibinfo{journal}{Physical review letters}}
  \textbf{\bibinfo{volume}{110}}, \bibinfo{pages}{168701}
  (\bibinfo{year}{2013}).

\bibitem{gonzalez2008understanding}
\bibinfo{author}{Gonzalez, M.~C.}, \bibinfo{author}{Hidalgo, C.~A.} \&
  \bibinfo{author}{Barabasi, A.-L.}
\newblock \bibinfo{title}{Understanding individual human mobility patterns}.
\newblock \emph{\bibinfo{journal}{Nature}} \textbf{\bibinfo{volume}{453}},
  \bibinfo{pages}{779--782} (\bibinfo{year}{2008}).

\bibitem{eagle2009eigenbehaviors}
\bibinfo{author}{Eagle, N.} \& \bibinfo{author}{Pentland, A.~S.}
\newblock \bibinfo{title}{Eigenbehaviors: Identifying structure in routine}.
\newblock \emph{\bibinfo{journal}{Behavioral Ecology and Sociobiology}}
  \textbf{\bibinfo{volume}{63}}, \bibinfo{pages}{1057--1066}
  (\bibinfo{year}{2009}).

\bibitem{song2010limits}
\bibinfo{author}{Song, C.}, \bibinfo{author}{Qu, Z.}, \bibinfo{author}{Blumm,
  N.} \& \bibinfo{author}{Barab{\'a}si, A.-L.}
\newblock \bibinfo{title}{Limits of predictability in human mobility}.
\newblock \emph{\bibinfo{journal}{Science}} \textbf{\bibinfo{volume}{327}},
  \bibinfo{pages}{1018--1021} (\bibinfo{year}{2010}).

\bibitem{lu2012predictability}
\bibinfo{author}{Lu, X.}, \bibinfo{author}{Bengtsson, L.} \&
  \bibinfo{author}{Holme, P.}
\newblock \bibinfo{title}{Predictability of population displacement after the
  2010 haiti earthquake}.
\newblock \emph{\bibinfo{journal}{Proceedings of the National Academy of
  Sciences}} \textbf{\bibinfo{volume}{109}}, \bibinfo{pages}{11576--11581}
  (\bibinfo{year}{2012}).

\bibitem{stopczynski2014measuring}
\bibinfo{author}{Stopczynski, A.} \emph{et~al.}
\newblock \bibinfo{title}{Measuring large-scale social networks with high
  resolution}.
\newblock \emph{\bibinfo{journal}{{PL}o{S} {ONE}}}
  \textbf{\bibinfo{volume}{9}}, \bibinfo{pages}{e95978} (\bibinfo{year}{2014}).

\bibitem{mucha2010community}
\bibinfo{author}{Mucha, P.~J.}, \bibinfo{author}{Richardson, T.},
  \bibinfo{author}{Macon, K.}, \bibinfo{author}{Porter, M.~A.} \&
  \bibinfo{author}{Onnela, J.-P.}
\newblock \bibinfo{title}{Community structure in time-dependent, multiscale,
  and multiplex networks}.
\newblock \emph{\bibinfo{journal}{Science}} \textbf{\bibinfo{volume}{328}},
  \bibinfo{pages}{876--878} (\bibinfo{year}{2010}).

\bibitem{gauvin2014detecting}
\bibinfo{author}{Gauvin, L.}, \bibinfo{author}{Panisson, A.} \&
  \bibinfo{author}{Cattuto, C.}
\newblock \bibinfo{title}{Detecting the community structure and activity
  patterns of temporal networks: a non-negative tensor factorization approach}.
\newblock \emph{\bibinfo{journal}{PLOS ONE}} \textbf{\bibinfo{volume}{9}},
  \bibinfo{pages}{e86028} (\bibinfo{year}{2014}).

\bibitem{rosvall2014memory}
\bibinfo{author}{Rosvall, M.}, \bibinfo{author}{Esquivel, A.~V.},
  \bibinfo{author}{Lancichinetti, A.}, \bibinfo{author}{West, J.~D.} \&
  \bibinfo{author}{Lambiotte, R.}
\newblock \bibinfo{title}{Memory in network flows and its effects on spreading
  dynamics and community detection}.
\newblock \emph{\bibinfo{journal}{Nature communications}}
  \textbf{\bibinfo{volume}{5}} (\bibinfo{year}{2014}).

\bibitem{greene2010tracking}
\bibinfo{author}{Greene, D.}, \bibinfo{author}{Doyle, D.} \&
  \bibinfo{author}{Cunningham, P.}
\newblock \bibinfo{title}{Tracking the evolution of communities in dynamic
  social networks}.
\newblock In \emph{\bibinfo{booktitle}{Advances in Social Networks Analysis and
  Mining (ASONAM), 2010 International Conference on}},
  \bibinfo{pages}{176--183} (\bibinfo{year}{2010}).

\bibitem{crandall2010inferring}
\bibinfo{author}{Crandall, D.~J.} \emph{et~al.}
\newblock \bibinfo{title}{Inferring social ties from geographic coincidences}.
\newblock \emph{\bibinfo{journal}{Proceedings of the National Academy of
  Sciences}} \textbf{\bibinfo{volume}{107}}, \bibinfo{pages}{22436--22441}
  (\bibinfo{year}{2010}).

\bibitem{wang2011human}
\bibinfo{author}{Wang, D.}, \bibinfo{author}{Pedreschi, D.},
  \bibinfo{author}{Song, C.}, \bibinfo{author}{Giannotti, F.} \&
  \bibinfo{author}{Barabasi, A.-L.}
\newblock \bibinfo{title}{Human mobility, social ties, and link prediction}.
\newblock In \emph{\bibinfo{booktitle}{Proceedings of the 17th ACM SIGKDD
  international conference on Knowledge discovery and data mining}},
  \bibinfo{pages}{1100--1108} (\bibinfo{organization}{ACM},
  \bibinfo{year}{2011}).

\bibitem{de2013interdependence}
\bibinfo{author}{De~Domenico, M.}, \bibinfo{author}{Lima, A.} \&
  \bibinfo{author}{Musolesi, M.}
\newblock \bibinfo{title}{Interdependence and predictability of human mobility
  and social interactions}.
\newblock \emph{\bibinfo{journal}{Pervasive and Mobile Computing}}
  \textbf{\bibinfo{volume}{9}}, \bibinfo{pages}{798--807}
  (\bibinfo{year}{2013}).

\bibitem{cho2011friendship}
\bibinfo{author}{Cho, E.}, \bibinfo{author}{Myers, S.~A.} \&
  \bibinfo{author}{Leskovec, J.}
\newblock \bibinfo{title}{Friendship and mobility: user movement in
  location-based social networks}.
\newblock In \emph{\bibinfo{booktitle}{Proceedings of the 17th ACM SIGKDD
  international conference on Knowledge discovery and data mining}},
  \bibinfo{pages}{1082--1090} (\bibinfo{organization}{ACM},
  \bibinfo{year}{2011}).

\bibitem{fano1968transmission}
\bibinfo{author}{Fano, R.~M.}
\newblock \emph{\bibinfo{title}{Transmission of information: a statistical
  theory of communications}} (\bibinfo{publisher}{MIT Press},
  \bibinfo{year}{1961}).

\bibitem{lin2012predictability}
\bibinfo{author}{Lin, M.}, \bibinfo{author}{Hsu, W.-J.} \&
  \bibinfo{author}{Lee, Z.~Q.}
\newblock \bibinfo{title}{Predictability of individuals' mobility with
  high-resolution positioning data}.
\newblock In \emph{\bibinfo{booktitle}{Proceedings of the 2012 ACM Conference
  on Ubiquitous Computing}}, \bibinfo{pages}{381--390}
  (\bibinfo{organization}{ACM}, \bibinfo{year}{2012}).

\bibitem{rosvall2010mapping}
\bibinfo{author}{Rosvall, M.} \& \bibinfo{author}{Bergstrom, C.~T.}
\newblock \bibinfo{title}{Mapping change in large networks}.
\newblock \emph{\bibinfo{journal}{PLoS ONE}} \textbf{\bibinfo{volume}{5}},
  \bibinfo{pages}{1--7} (\bibinfo{year}{2010}).
\newblock \urlprefix\url{http://dx.doi.org/10.1371%2Fjournal.pone.0008694}.

\bibitem{de2015identifying}
\bibinfo{author}{De~Domenico, M.}, \bibinfo{author}{Lancichinetti, A.},
  \bibinfo{author}{Arenas, A.} \& \bibinfo{author}{Rosvall, M.}
\newblock \bibinfo{title}{Identifying modular flows on multilayer networks
  reveals highly overlapping organization in interconnected systems}.
\newblock \emph{\bibinfo{journal}{Physical Review X}}
  \textbf{\bibinfo{volume}{5}}, \bibinfo{pages}{011027} (\bibinfo{year}{2015}).

\bibitem{salnikov2016using}
\bibinfo{author}{Salnikov, V.}, \bibinfo{author}{Schaub, M.~T.} \&
  \bibinfo{author}{Lambiotte, R.}
\newblock \bibinfo{title}{Using higher-order markov models to reveal flow-based
  communities in networks}.
\newblock \emph{\bibinfo{journal}{Scientific Reports}}
  \textbf{\bibinfo{volume}{6}} (\bibinfo{year}{2016}).

\bibitem{sekara2014strength}
\bibinfo{author}{Sekara, V.} \& \bibinfo{author}{Lehmann, S.}
\newblock \bibinfo{title}{The strength of friendship ties in proximity sensor
  data}.
\newblock \emph{\bibinfo{journal}{PLoS One}} \textbf{\bibinfo{volume}{9}},
  \bibinfo{pages}{e100915} (\bibinfo{year}{2014}).

\bibitem{eagle2009inferring}
\bibinfo{author}{Eagle, N.}, \bibinfo{author}{Pentland, A.~S.} \&
  \bibinfo{author}{Lazer, D.}
\newblock \bibinfo{title}{Inferring friendship network structure by using
  mobile phone data}.
\newblock \emph{\bibinfo{journal}{Proceedings of the National Academy of
  Sciences}} \textbf{\bibinfo{volume}{106}}, \bibinfo{pages}{15274--15278}
  (\bibinfo{year}{2009}).

\bibitem{clauset2012persistence}
\bibinfo{author}{Clauset, A.} \& \bibinfo{author}{Eagle, N.}
\newblock \bibinfo{title}{Persistence and periodicity in a dynamic proximity
  network}.
\newblock \emph{\bibinfo{journal}{arXiv preprint arXiv:1211.7343}}
  (\bibinfo{year}{2012}).

\bibitem{sulo2010meaningful}
\bibinfo{author}{Sulo, R.}, \bibinfo{author}{Berger-Wolf, T.} \&
  \bibinfo{author}{Grossman, R.}
\newblock \bibinfo{title}{Meaningful selection of temporal resolution for
  dynamic networks}.
\newblock In \emph{\bibinfo{booktitle}{Proceedings of the Eighth Workshop on
  Mining and Learning with Graphs}}, \bibinfo{pages}{127--136}
  (\bibinfo{organization}{ACM}, \bibinfo{year}{2010}).

\bibitem{krings2012effects}
\bibinfo{author}{Krings, G.}, \bibinfo{author}{Karsai, M.},
  \bibinfo{author}{Bernhardsson, S.}, \bibinfo{author}{Blondel, V.~D.} \&
  \bibinfo{author}{Saram{\"a}ki, J.}
\newblock \bibinfo{title}{Effects of time window size and placement on the
  structure of an aggregated communication network}.
\newblock \emph{\bibinfo{journal}{EPJ Data Science}}
  \textbf{\bibinfo{volume}{1}}, \bibinfo{pages}{1--16} (\bibinfo{year}{2012}).

\bibitem{coser1971masters}
\bibinfo{author}{Coser, L.~A.} \& \bibinfo{author}{Merton, R.~K.}
\newblock \emph{\bibinfo{title}{Masters of sociological thought: Ideas in
  historical and social context}} (\bibinfo{publisher}{Harcourt Brace
  Jovanovich New York}, \bibinfo{year}{1971}).

\bibitem{ward1963hierarchical}
\bibinfo{author}{Ward~Jr, J.~H.}
\newblock \bibinfo{title}{Hierarchical grouping to optimize an objective
  function}.
\newblock \emph{\bibinfo{journal}{Journal of the American statistical
  association}} \textbf{\bibinfo{volume}{58}}, \bibinfo{pages}{236--244}
  (\bibinfo{year}{1963}).

\bibitem{gower1969minimum}
\bibinfo{author}{Gower, J.~C.} \& \bibinfo{author}{Ross, G.}
\newblock \bibinfo{title}{Minimum spanning trees and single linkage cluster
  analysis}.
\newblock \emph{\bibinfo{journal}{Applied statistics}} \bibinfo{pages}{54--64}
  (\bibinfo{year}{1969}).

\bibitem{newman2004finding}
\bibinfo{author}{Newman, M.~E.} \& \bibinfo{author}{Girvan, M.}
\newblock \bibinfo{title}{Finding and evaluating community structure in
  networks}.
\newblock \emph{\bibinfo{journal}{Physical review E}}
  \textbf{\bibinfo{volume}{69}}, \bibinfo{pages}{026113}
  (\bibinfo{year}{2004}).

\bibitem{reis1991studying}
\bibinfo{author}{Reis, H.~T.} \& \bibinfo{author}{Wheeler, L.}
\newblock \bibinfo{title}{Studying social interaction with the rochester
  interaction record}.
\newblock \emph{\bibinfo{journal}{Advances in experimental social psychology}}
  \textbf{\bibinfo{volume}{24}}, \bibinfo{pages}{269--318}
  (\bibinfo{year}{1991}).

\bibitem{davis1941deep}
\bibinfo{author}{Davis, A.}, \bibinfo{author}{Gardner, B.~B.} \&
  \bibinfo{author}{Gardner, M.~R.}
\newblock \emph{\bibinfo{title}{Deep South; a social anthropological study of
  caste and class.}} (\bibinfo{publisher}{University of Chicago Press},
  \bibinfo{year}{1941}).

\bibitem{tibshirani2001estimating}
\bibinfo{author}{Tibshirani, R.}, \bibinfo{author}{Walther, G.} \&
  \bibinfo{author}{Hastie, T.}
\newblock \bibinfo{title}{Estimating the number of clusters in a data set via
  the gap statistic}.
\newblock \emph{\bibinfo{journal}{Journal of the Royal Statistical Society:
  Series B (Statistical Methodology)}} \textbf{\bibinfo{volume}{63}},
  \bibinfo{pages}{411--423} (\bibinfo{year}{2001}).

\bibitem{milligan1985examination}
\bibinfo{author}{Milligan, G.~W.} \& \bibinfo{author}{Cooper, M.~C.}
\newblock \bibinfo{title}{An examination of procedures for determining the
  number of clusters in a data set}.
\newblock \emph{\bibinfo{journal}{Psychometrika}}
  \textbf{\bibinfo{volume}{50}}, \bibinfo{pages}{159--179}
  (\bibinfo{year}{1985}).

\bibitem{shannon1948entropy}
\bibinfo{author}{Shannon, C.}
\newblock \bibinfo{title}{A mathematical theory of communication}.
\newblock \emph{\bibinfo{journal}{The Bell System Technical Journal}}
  \textbf{\bibinfo{volume}{27}}, \bibinfo{pages}{379--423}
  (\bibinfo{year}{1948}).

\bibitem{jensen2010estimating}
\bibinfo{author}{Jensen, B.~S.}, \bibinfo{author}{Larsen, J.~E.},
  \bibinfo{author}{Jensen, K.}, \bibinfo{author}{Larsen, J.} \&
  \bibinfo{author}{Hansen, L.~K.}
\newblock \bibinfo{title}{Estimating human predictability from mobile sensor
  data}.
\newblock In \emph{\bibinfo{booktitle}{Machine Learning for Signal Processing
  (MLSP), 2010 IEEE International Workshop on}}, \bibinfo{pages}{196--201}
  (\bibinfo{organization}{IEEE}, \bibinfo{year}{2010}).

\bibitem{bagrow2012mesoscopic}
\bibinfo{author}{Bagrow, J.~P.} \& \bibinfo{author}{Lin, Y.-R.}
\newblock \bibinfo{title}{Mesoscopic structure and social aspects of human
  mobility}.
\newblock \emph{\bibinfo{journal}{PloS one}} \textbf{\bibinfo{volume}{7}},
  \bibinfo{pages}{e37676} (\bibinfo{year}{2012}).

\bibitem{trevisani2004cell}
\bibinfo{author}{Trevisani, E.} \& \bibinfo{author}{Vitaletti, A.}
\newblock \bibinfo{title}{Cell-id location technique, limits and benefits: an
  experimental study}.
\newblock In \emph{\bibinfo{booktitle}{Mobile Computing Systems and
  Applications, 2004. WMCSA 2004. Sixth IEEE Workshop on}},
  \bibinfo{pages}{51--60} (\bibinfo{organization}{IEEE}, \bibinfo{year}{2004}).

\bibitem{cuttone2014inferring}
\bibinfo{author}{Cuttone, A.}, \bibinfo{author}{Lehmann, S.} \&
  \bibinfo{author}{Larsen, J.~E.}
\newblock \bibinfo{title}{Inferring human mobility from sparse low accuracy
  mobile sensing data}.
\newblock In \emph{\bibinfo{booktitle}{Proceedings of the 2014 ACM
  International Joint Conference on Pervasive and Ubiquitous Computing: Adjunct
  Publication}}, \bibinfo{pages}{995--1004} (\bibinfo{organization}{ACM},
  \bibinfo{year}{2014}).

\bibitem{ester1996density}
\bibinfo{author}{Ester, M.}, \bibinfo{author}{Kriegel, H.-P.},
  \bibinfo{author}{Sander, J.} \& \bibinfo{author}{Xu, X.}
\newblock \bibinfo{title}{A density-based algorithm for discovering clusters in
  large spatial databases with noise.}
\newblock In \emph{\bibinfo{booktitle}{Kdd}}, vol.~\bibinfo{volume}{96},
  \bibinfo{pages}{226--231} (\bibinfo{year}{1996}).

\bibitem{wrzus2013social}
\bibinfo{author}{Wrzus, C.}, \bibinfo{author}{H{\"a}nel, M.},
  \bibinfo{author}{Wagner, J.} \& \bibinfo{author}{Neyer, F.~J.}
\newblock \bibinfo{title}{Social network changes and life events across the
  life span: A meta-analysis.}
\newblock \emph{\bibinfo{journal}{Psychological bulletin}}
  \textbf{\bibinfo{volume}{139}}, \bibinfo{pages}{53} (\bibinfo{year}{2013}).

\bibitem{penrose2003random}
\bibinfo{author}{Penrose, M.}
\newblock \emph{\bibinfo{title}{Random geometric graphs}}
  (\bibinfo{publisher}{Oxford University Press Oxford}, \bibinfo{year}{2003}).

\bibitem{barthelemy2011spatial}
\bibinfo{author}{Barthelemy, M.}
\newblock \bibinfo{title}{Spatial networks}.
\newblock \emph{\bibinfo{journal}{Physics Reports}}
  \textbf{\bibinfo{volume}{499}}, \bibinfo{pages}{1 -- 101}
  (\bibinfo{year}{2011}).

\bibitem{PhysRevE.66.016121}
\bibinfo{author}{Dall, J.} \& \bibinfo{author}{Christensen, M.}
\newblock \bibinfo{title}{Random geometric graphs}.
\newblock \emph{\bibinfo{journal}{Phys. Rev. E}} \textbf{\bibinfo{volume}{66}},
  \bibinfo{pages}{016121} (\bibinfo{year}{2002}).
\newblock \urlprefix\url{http://link.aps.org/doi/10.1103/PhysRevE.66.016121}.

\bibitem{0305-4470-33-42-104}
\bibinfo{author}{Quintanilla, J.}, \bibinfo{author}{Torquato, S.} \&
  \bibinfo{author}{Ziff, R.~M.}
\newblock \bibinfo{title}{Efficient measurement of the percolation threshold
  for fully penetrable discs}.
\newblock \emph{\bibinfo{journal}{Journal of Physics A: Mathematical and
  General}} \textbf{\bibinfo{volume}{33}}, \bibinfo{pages}{L399}
  (\bibinfo{year}{2000}).
\newblock \urlprefix\url{http://stacks.iop.org/0305-4470/33/i=42/a=104}.

\bibitem{kullback1951information}
\bibinfo{author}{Kullback, S.} \& \bibinfo{author}{Leibler, R.~A.}
\newblock \bibinfo{title}{On information and sufficiency}.
\newblock \emph{\bibinfo{journal}{The annals of mathematical statistics}}
  \textbf{\bibinfo{volume}{22}}, \bibinfo{pages}{79--86}
  (\bibinfo{year}{1951}).

\bibitem{PhysRevLett.96.088702}
\bibinfo{author}{Gonz\'alez, M.~C.}, \bibinfo{author}{Lind, P.~G.} \&
  \bibinfo{author}{Herrmann, H.~J.}
\newblock \bibinfo{title}{System of mobile agents to model social networks}.
\newblock \emph{\bibinfo{journal}{Phys. Rev. Lett.}}
  \textbf{\bibinfo{volume}{96}}, \bibinfo{pages}{088702}
  (\bibinfo{year}{2006}).
\newblock
  \urlprefix\url{http://link.aps.org/doi/10.1103/PhysRevLett.96.088702}.

\bibitem{onnela2007structure}
\bibinfo{author}{Onnela, J.-P.} \emph{et~al.}
\newblock \bibinfo{title}{Structure and tie strengths in mobile communication
  networks}.
\newblock \emph{\bibinfo{journal}{Proceedings of the National Academy of
  Sciences}} \textbf{\bibinfo{volume}{104}}, \bibinfo{pages}{7332--7336}
  (\bibinfo{year}{2007}).

\bibitem{jo2012circadian}
\bibinfo{author}{Jo, H.-H.}, \bibinfo{author}{Karsai, M.},
  \bibinfo{author}{Kert{\'e}sz, J.} \& \bibinfo{author}{Kaski, K.}
\newblock \bibinfo{title}{Circadian pattern and burstiness in mobile phone
  communication}.
\newblock \emph{\bibinfo{journal}{New Journal of Physics}}
  \textbf{\bibinfo{volume}{14}}, \bibinfo{pages}{013055}
  (\bibinfo{year}{2012}).

\end{thebibliography}


\begin{thebibliography}{10}

\bibitem{easley2010networks}
Easley D, Kleinberg J
\newblock (2010) \emph{Networks, crowds, and markets: Reasoning about a highly
  connected world}
\newblock (Cambridge University Press).

\bibitem{newman2010networks}
Newman M
\newblock (2010) \emph{Networks: An Introduction}
\newblock (Oxford University Press).


\bibitem{wasserman1994social}
Wasserman S, Faust, K
\newblock (1994) \emph{Social Network Analysis: Methods and Applications}.
\newblock (Cambridge University Press).

\bibitem{simmel1950quantitative}
Simmel G
\newblock (1950) Quantitative aspects of the group.
\newblock \emph{The Sociology of Georg Simmel} pp 87--177.

\bibitem{goffman2005interaction}
Goffman E
\newblock (2005) \emph{Interaction ritual: Essays in {F}ace to {F}ace
  {B}ehavior}
\newblock (AldineTransaction).

\bibitem{palla2005uncovering}
Palla G, Der{\'e}nyi I, Farkas I, Vicsek T
\newblock (2005) Uncovering the overlapping community structure of complex
  networks in nature and society.
\newblock \emph{Nature} 435:814--818.

\bibitem{palla2007quantifying}
Palla G, Barab{\'a}si AL, Vicsek T
\newblock (2007) Quantifying social group evolution.
\newblock \emph{Nature} 446:664--667.

\bibitem{clauset2008hierarchical}
Clauset A, Moore C, Newman ME
\newblock (2008) Hierarchical structure and the prediction of missing links in
  networks.
\newblock \emph{Nature} 453:98--101.

\bibitem{ahn2010link}
Ahn YY, Bagrow JP, Lehmann S
\newblock (2010) Link communities reveal multiscale complexity in networks.
\newblock \emph{Nature} 466:761--764.

\bibitem{holme2012temporal}
Holme P, Saram{\"a}ki J
\newblock (2012) Temporal networks.
\newblock \emph{Physics Reports} 519:97--125.

\bibitem{starnini2013modeling}
Starnini M, Baronchelli A, Pastor-Satorras R
\newblock (2013) Modeling human dynamics of face-to-face interaction networks.
\newblock \emph{Physical review letters} 110:168701.

\bibitem{gonzalez2008understanding}
Gonzalez MC, Hidalgo CA, Barabasi AL
\newblock (2008) Understanding individual human mobility patterns.
\newblock \emph{Nature} 453:779--782.

\bibitem{eagle2009eigenbehaviors}
Eagle N, Pentland, AS
\newblock (2009) Eigenbehaviors: Identifying structure in routine.
\newblock \emph{Behavioral Ecology and Sociobiology} 63:1057--1066.

\bibitem{song2010limits}
Song C, Qu Z, Blumm N, Barab{\'a}si AL
\newblock (2010) Limits of predictability in human mobility.
\newblock \emph{Science} 327:1018--1021.

\bibitem{lu2012predictability}
Lu X, Bengtsson L, Holme P
\newblock (2012) Predictability of population displacement after the 2010 Haiti earthquake.
\newblock \emph{Proceedings of the National Academy of Sciences} 109:11576--11581.

\bibitem{saramaki2015seconds}
{Saram\"{a}ki, Jari}, {Moro, Esteban}
\newblock (2015) From seconds to months: an overview of multi-scale dynamics of
  mobile telephone calls.
\newblock \emph{Eur. Phys. J. B} 88:164.

\bibitem{stopczynski2014measuring}
Stopczynski A, {et~al.}
\newblock (2014) Measuring large-scale social networks with high resolution.
\newblock \emph{PLoS ONE} 9:e95978.

\bibitem{mucha2010community}
Mucha PJ, Richardson T, Macon K, Porter MA, Onnela JP
\newblock (2010) Community structure in time-dependent, multiscale, and
  multiplex networks.
\newblock \emph{Science} 328:876--878.

\bibitem{gauvin2014detecting}
Gauvin L, Panisson A, Cattuto C
\newblock (2014) Detecting the community structure and activity patterns of
  temporal networks: a non-negative tensor factorization approach.
\newblock \emph{PLoS ONE} 9:e86028.

\bibitem{golder2011diurnal}
Golder SA, Macy MW
\newblock (2011) Diurnal and seasonal mood vary with work, sleep, and daylength
  across diverse cultures.
\newblock \emph{Science} 333:1878--1881.

\bibitem{rosvall2014memory}
Rosvall M, Esquivel AV, Lancichinetti A, West JD, Lambiotte R
\newblock (2014) Memory in network flows and its effects on spreading dynamics
  and community detection.
\newblock \emph{Nature communications} 5.

\bibitem{evans2009line}
Evans TS, Lambiotte R
\newblock (2009) Line graphs, link partitions, and overlapping communities.
\newblock \emph{Physical Review E} 80:016105.

\bibitem{greene2010tracking}
Greene D, Doyle D, Cunningham P
\newblock (2010) \emph{Tracking the Evolution of Communities in Dynamic Social
  Networks}
\newblock pp 176--183.

\bibitem{crandall2010inferring}
Crandall DJ, Backstrom L, Cosley D, Suri S, Huttenlocher D, Kleinberg J
\newblock (2010) Inferring social ties from geographic coincidences.
\newblock \emph{Proceedings of the National Academy of Sciences} 107:22436--22441.

\bibitem{wang2011human}
Wang D, Pedreschi D, Song C, Giannotti F, Barab{\'a}si AL
\newblock (2011) \emph{Human mobility, social ties, and link prediction}
\newblock (ACM), pp 1100--1108.

\bibitem{de2013interdependence}
De~Domenico M, Lima A, Musolesi M
\newblock (2013) Interdependence and predictability of human mobility and
  social interactions.
\newblock \emph{Pervasive and Mobile Computing} 9:798--807.

\bibitem{cho2011friendship}
Cho E, Myers SA, Leskovec J
\newblock (2011) Friendship and mobility: user movement in location-based social networks.
\newblock (Proceedings of the 17th ACM SIGKDD International Conference on Knowledge Discovery and Data Mining).

\bibitem{miritello2013temporal}
Miritello G
\newblock (2013) \emph{Temporal patterns of communication in social networks}
\newblock (Springer Science \& Business Media).

\bibitem{fano1968transmission}
Fano RM
\newblock (1961) \emph{Transmission of information: a statistical theory of
  communications}
\newblock (MIT Press).

\bibitem{lu2013approaching}
Lu X, Wetter E, Bharti N, Tatem AJ, Bengtsson L
\newblock (2013) Approaching the limit of predictability in human mobility.
\newblock \emph{Scientific reports} 3.

\bibitem{lin2012predictability}
Lin M, Hsu WJ, Lee ZQ
\newblock (2012) \emph{Predictability of individuals' mobility with
  high-resolution positioning data}
\newblock (ACM), pp 381--390.

\bibitem{mcinerney2012exploring}
McInerney J, Stein S, Rogers A, Jennings NR
\newblock (2012) \emph{Exploring periods of low predictability in daily life
  mobility}.

\bibitem{rosvall2010mapping}
Rosvall M, Bergstrom CT
\newblock (2010) Mapping change in large networks.
\newblock \emph{PLoS ONE} 5:1--7.

\bibitem{de2015identifying}
De~Domenico M, Lancichinetti A, Arenas A, Rosvall M
\newblock (2015) Identifying modular flows on multilayer networks reveals
  highly overlapping organization in interconnected systems.
\newblock \emph{Physical Review X} 5:011027.

\bibitem{salnikov2016using}
Salnikov V, Schaub MT, Lambiotte R
\newblock (2016) Using higher-order markov models to reveal flow-based
  communities in networks.
\newblock \emph{Scientific Reports} 6.

\end{thebibliography}
\bibliographystyle{naturemag}

\end{document}